\documentclass[12pt]{article}
\usepackage{amscd,amsmath,amssymb,amstext,amsthm,exscale,latexsym}
\usepackage{graphicx,floatflt}
\newcommand {\al}   {\alpha}       \newcommand {\bt}  {\beta}
\newcommand {\g }   {\gamma}       \newcommand {\G }  {\Gamma}
\newcommand {\dl}   {\delta}       
        
\newcommand {\ve}   {\varepsilon}  
\newcommand {\lm}   {\lambda}      
          
\newcommand {\s }   {\sigma}

\newcommand {\Lm}   {\Lambda}      
       
\newcommand {\pl}   {\partial}     \newcommand {\nb}  {\nabla}
\newcommand {\Sa}  {{\textsc{a}}}   \newcommand {\Sb}  {{\textsc{b}}}
\newcommand {\Se}  {{\textsc{e}}}  \newcommand {\Sh}  {{\textsc{h}}}

\renewcommand {\det}{{\sf\,det\,}}       \renewcommand {\exp}{{\sf\,exp\,}}
\renewcommand {\dim}{{\sf\,dim\,}}       \renewcommand {\deg}{{\sf\,deg\,}}

\newcommand   {\const}{{\sf\,const}}     
\newcommand {\MM}  {{\mathbb M}}   \newcommand {\MN}  {{\mathbb N}}
\newcommand {\MR}  {{\mathbb R}}   \newcommand {\MT}  {{\mathbb T}}
\newcommand {\MU}  {{\mathbb U}}   \newcommand {\MV}  {{\mathbb V}}
\newcommand {\CE}  {{\cal E}}      \newcommand {\CH}  {{\cal H}}
\newcommand {\CL}  {{\cal L}}      \newcommand {\CP}  {{\cal P}}
\newcommand {\Sd}  {{\textsc{d}}}
\newcommand {\Sm}  {{\textsc{m}}}
\newcommand {\Bx}  {\boldsymbol{x}}
\newcommand {\CC }  {{\cal C}}
\begin{document}
\title     {Polynomial Hamiltonian form of General Relativity}
\author    {M. O. Katanaev
            \thanks{E-mail: katanaev@mi.ras.ru}\\ \\
            \sl Steklov Mathematical Institute,\\
            \sl Gubkin St.~8, Moscow, 119991}
\date      {27 October 2005}
\maketitle
\begin{abstract}
  The phase space of general relativity is extended to a Poisson manifold
  by inclusion of the determinant of the metric and conjugate momentum
  as additional independent variables. As a result, the action and the
  constraints take a polynomial form. We propose a new expression for the
  generating functional for the Green's functions. We show that the Dirac
  bracket defines a degenerate Poisson structure on a manifold and the second
  class constraints are the Casimir functions with respect to this structure.
  As an application of the new variables, we consider the Friedmann universe.
\end{abstract}
\section{Introduction}
Canonically formulating any model of mathematical physics is the most
important step when analyzing equations of motion, in particular, when
setting and analyzing the Cauchy problem. It also provides the
basis for canonically quantizing models. Canonically formulating general
relativity is technically involved, and many papers are devoted to this
problem. We mention only a few of them. Dirac first formulated general
relativity self-consistently in the second-order formulism \cite{Dirac58B}.
He took the metric components $g_{\al\bt}$ as the independent variables
and showed that the Hamiltonian of the gravitational field is equal to a
linear combination of constraints. Afterwards, Arnowitt, Deser, and Misner
in the series of papers resulting in the review \cite{ArDeMi62} essentially
simplified the calculations and clarified the geometrical meaning of the
canonical momenta, expressing them in terms of the extrinsic curvature of
a spacelike hypersurface imbedded in a four-dimensional space-time. The
expression for the Hamiltonian was found in the first order formalism when
the metric $g_{\al\bt}$ and the symmetric affine connection
$\G_{\lbrace\al\bt\rbrace}{}^\g$ considered independent variables.
In essence, this approach simplified the calculations. In the pioneering
papers \cite{Dirac58B,ArDeMi62}, the constraint algebra was not calculated
explicitly, but the constraints were shown to be consistent with the equations
of motion (the first class constraints). In \cite{ArDeMi62}, the role
of boundary terms was also analyzed, and the total energy of the gravitational
field for an asymptotically flat space-time was defined in terms of the
surface integral. The role of boundary terms was discussed in a more general
form in \cite{RegTei74}.

In \cite{DeWitt67A}, general relativity was canonically formulated and a
generalization of the Schr\"odinger equation for the wave function of
the universe, which later was called the Wheeler--DeWitt equation, was
considered in detail. The constraint algebra in general relativity was first
calculated explicitly there.

The constraint algebra for any model invariant with respect to general
coordinate transformations was obtained in \cite{Teitel73} assuming that the
model is self-contained and that the constraints generate the general coordinate
transformations for the canonical variables (see also \cite{Geroch76B}).
We note that the self-consistency of a model (the closedness of the
constraint algebra) is a very strong assumption. Conversely, for a given
model, the constraint algebra, because it is not known beforehand, must be
calculated explicitly to prove the self-consistency of the model.

Dirac \cite{Dirac62} and Schwinger \cite{Schwin63A} started the investigation
of the vielbein Hamiltonian formulation of general relativity using the
time gauge for simplicity. The Hamiltonian formulation in a general case
without gauge fixing was given much later because of serious technical
difficulties \cite{DesIsh76,ChHeNe88}.

The Hamiltonian formulation of general relativity contains constraints that
are nonpolynomial on space-section metric components, and this is an essential
obstacle to analyzing and quantizing the theory of gravity. The polynomial
Hamiltonian formulation given by Ashtekar \cite{Ashtek87} attracted
much interest in recent years. He proposed using complex variables in the
extended phase space, which are tensor densities and lead to polynomial
constraints. Here, we consider a different extension of the phase space
\cite{Tate92} with the metric determinant and its conjugate momentum
considered additional variables. We show that the Poisson structure on the
extended phase space is degenerate and that the initial phase space is
mapped on a subspace of the extended phase space by the canonical transformation.
All new canonical variables are real tensor densities, and the constraints take a
polynomial form.

We propose a functional integral over a Poisson manifold as a new expression
for the generating functional for the Green's functions. This form of
the integral reduces to the standard expression for the generating functional
over the phase space \cite{Faddee69} after integration over the additional
variables, which is removed by two supplementary $\dl$-functions. We prove that
the corresponding Jacobian of coordinate transformation is equal to unity.

The plan of the paper is as follows. In Secs.\ \ref{sadmpa}--\ref{salcog},
we describe the transition from the Hilbert--Einstein Lagrangian to the
Hamiltonian in detail to avoid sending the reader to the original
papers, where a greater part of the calculations is usually omitted. Moreover,
we consider a general case of affine space-time geometry in Sec.\ \ref{sgeohy}
when describing the geometry of hypersurfaces. In this case, the antisymmetric
part of external curvature of a hypersurface is defined by the torsion tensor.
This is important for the canonical formulation of general relativity in
the vielbein formulation and for the Hamiltonian formulation of models with
absolute parallelism. The canonical transformation between the phase space
of general relativity and the submanifold of the extended Poisson manifold
is described in Sec.\ \ref{scatrg}. We show that all constraints and the
action of the model take a polynomial form in the extended space, and we
compute the constraint algebra. We propose the expression for the generating
functional for the Green's functions on the Poisson manifold in Sec.\
\ref{sgenfu}. In Sec.\ \ref{shoisp} as an application of the new variables,
we consider the case of a homogeneous and isotropic universe, a case
where all the calculations can be easily checked.
\section{ADM parameterization of a metric             \label{sadmpa}}
To analyze the Hamiltonian structure of the general relativity equations
Arnowitt, Deser, and Misner used the special parameterization of the metric
(ADM-parameterization), which essentially simplifies calculations \cite{ArDeMi62}.
We consider a manifold $\MM$, $\dim\MM=n$ equipped with a metric of
Lorentzian signature $(+-\dotsc-)$. We deliberately do not restrict ourselves
to the most important case of four-dimensional space-time because gravity
models in higher and lower number of dimensions have attracted much interest
recently. Let $\lbrace x^\al\rbrace$, $\al=0,1,\dotsc,n-1$ denote the
local coordinates. We choose the time coordinate $t=x^0$, and then
$\lbrace x^\al\rbrace=\lbrace x^0,x^\mu\rbrace$, $\mu=1,\dotsc,n-1$.
In what follows, the letters from the beginning of the Greek alphabet
$(\al,\bt,\dotsc)$ range all index values, while the letters
from the middle $(\mu,\nu,\dotsc)$ range only the space-related values.
This rule is easily remembered from the inclusions
$\lbrace\mu,\nu,\dotsc\rbrace\subset\lbrace\al,\bt,\dotsc\rbrace$ and
$\lbrace1,2,\dotsc\rbrace\subset\lbrace0,1,2,\dotsc\rbrace$.
The ADM parameterization of the metric has the form
\begin{equation}                                        \label{eadmme}
g_{\al\bt}=\begin{pmatrix} N^2+N^\rho N_\rho & N_\nu \\
           N_\mu & g_{\mu\nu}\end{pmatrix},
\end{equation}
where $g_{\mu\nu}$ is the metric on $(n-1)$-dimensional manifold sections
$x^0=\const$. In the chosen parameterization, we introduced the same number
of functions $N$ and $N_\mu$ instead of the $n$ metric components containing
at least one time index $g_{00}$ and $g_{0\mu}$. Here, $N^\rho=\hat g^{\rho\mu}N_\mu$,
where $\hat g^{\rho\mu}$ is the $(n-1)\times(n-1)$-matrix inverse to $g_{\mu\nu}$:
\begin{equation*}
  \hat g^{\rho\mu}g_{\mu\nu}=\dl^\rho_\nu,
\end{equation*}
which we call the inverse metric on the sections $x^0=\const$. In what follows,
we always raise the space indices using the inverse metric $\hat g^{\rho\mu}$,
which is marked with the hat and does not coincide with the space part of the
metric $g^{\al\bt}$ inverse to $g_{\al\bt}$, $\hat g^{\rho\mu}\ne g^{\rho\mu}$.
The function $N=N(x)$ is called the lapse function, and the functions
$N_\mu=N_\mu(x)$ are shift functions. Without loss of generality, we assume
that the lapse function is positive ($N>0$). In this case, the ADM
parameterization of the metric (\ref{eadmme}) is in one-to-one correspondence.
The interval corresponding to the parameterization (\ref{eadmme}) has the form
\begin{equation*}
  ds^2=N^2dt^2+g_{\mu\nu}(dx^\mu+N^\mu dt)(dx^\nu+N^\nu dt).
\end{equation*}

We assume that the coordinate $x^0=t$ is the time, i.e., that the vector
$\pl_0$ tangent to the coordinate $x^0$ is timelike. Formally, this condition
is written as
\begin{equation}                                        \label{etimec}
  (\pl_0,\pl_0)=g_{00}=N^2+N^\rho N_\rho>0.
\end{equation}
In this case, the metric $g_{\al\bt}$ has the Lorentzian signature if and
only if the matrix
\begin{equation}                                        \label{espadm}
  g_{\mu\nu}-\frac{N_\mu N_\nu}{N^2+N^\rho N_\rho}
\end{equation}
is negative definite. We note that the metric $g_{\mu\nu}$ induced on sections
$x^0=\const$ may not be negative definite. This means that sections $x^0=\const$
are not spacelike in a general case. In what follows, we additionally
assume that the coordinates are chosen such that all sections $x^0=\const$ are
spacelike, i.e., the metric $g_{\mu\nu}$ is also negative definite. This is
convenient for posing the Cauchy problem when the initial data
are given on a spacelike surface and we consider their evolution in time.

Similarly, a metric on a Riemannian manifold can be parameterized with a
positive-definite metric. For this, it suffices to explicitly choose
any coordinate instead of time.

The metric inverse to (\ref{eadmme}) is
\begin{equation}                                        \label{eadmmi}
  g^{\al\bt}=\left(\begin{aligned}
  &\dfrac{\raise-.7ex\hbox{1}}{N^2}
  & &\!-\!\dfrac{\raise-.7ex\hbox{$N^\nu$}}{N^2} \\
  -&\dfrac{\raise-.7ex\hbox{$N^\mu$}}{N^2}
  & ~\hat g^{\mu\nu}\!\!\!&+\!\dfrac{\raise-.7ex\hbox{$N^\mu N^\nu$}}{N^2}
  \end{aligned}\right).
\end{equation}
The space matrix in the lower right block
\begin{equation}                                       \label{eincpm}
  g^{\mu\nu}=\hat g^{\mu\nu}+\frac{N^\mu N^\nu}{N^2},
\end{equation}
is inverse to metric (\ref{espadm}), as can be easily verified.
This means that the negative definiteness of metric (\ref{espadm})
is equivalent to the negative definiteness of the matrix $g^{\mu\nu}$.

\begin{floatingfigure}{.45\textwidth}
\includegraphics[width=.4\textwidth]{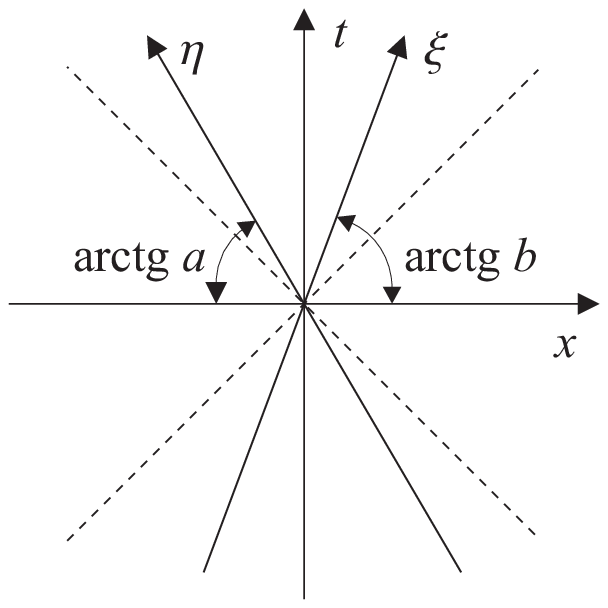}
 \label{fminpl}
\end{floatingfigure}

We note that if the metric on a manifold $\MM$ has the Lorentzian signature,
then the condition that all sections $x^0=\const$ are spacelike is equi\-valent
to the condition $N^2>0$. Indeed, the negative definiteness of the inverse
matrix $\hat g^{\mu\nu}$ follows from that of $g_{\mu\nu}$.
Then the negative definiteness of the matrix
\begin{equation*}
  g^{\mu\nu}-\frac{N^\mu N^\nu}{N^2}
\end{equation*}
follows from Eq.\ (\ref{eincpm}). In turn, this is equivalent to
the condition $g^{00}>0$ or $N^2>0$.

We consider a simple example to show details that may arise
for ADM para\-meterization of a metric.

{\bf Example.}
We consider the two-dimensional \linebreak Minkowskian space-time $\MR^{1,1}$
with the Cartesian coordinates $t,x$. We introduce the new coordinate system
$\xi,\eta$ depending on two real parameters $a$ and $b$ (see Figure)
\begin{equation*}
  \xi=t+ax,~~\eta=t-bx,~~~~~~|a|\ne1,~~|b|\ne1,~~a+b\ne0.
\end{equation*}
We can easily obtain the formulas for the inverse transformation
\begin{equation*}
  t=\frac{b\xi+a\eta}{a+b},~~~~x=\frac{\xi-\eta}{a+b}.
\end{equation*}
The metric has the form
\begin{equation*}
  ds^2=dt^2-dx^2=\frac1{(a+b)^2}\left[(b^2-1)d\xi^2+2(ab+1)d\xi d\eta
  +(a^2-1)d\eta^2\right].
\end{equation*}
in the new coordinates.

We now analyze the ADM parameterization of the metric in the coordinates
$x^0=\xi,~x^1=\eta$:
\begin{equation*}
  g_{00}=\frac{b^2-1}{(a+b)^2},~~~~g_{01}=\frac{ab+1}{(a+b)^2},~~~~
  g_{11}=\frac{a^2-1}{(a+b)^2}.
\end{equation*}
The lapse and shift functions are
\begin{equation*}
  N^2=-\frac1{a^2-1},~~~~N_1=\frac{ab+1}{(a+b)^2}.
\end{equation*}
The inequalities $|b|>1$ and $|a|<1$ follow from the respective conditions
$g_{00}>0$ and $g_{11}<0$. We see that these conditions are necessary
and sufficient for the coordinate lines $\xi$ and $\eta$ to be respectively
timelike and spacelike. It is easy to verify the equivalence of the conditions
\begin{align*}
  g_{00}>0~~~~&\Leftrightarrow~~~~
  g_{11}-\frac{N_1N_1}{N^2+N^1N_1}=-\frac1{b^2-1}<0,
\\                                                           \tag*{\qed}
  g^{00}>0~~~~&\Leftrightarrow~~~~
  \hat g^{11}=g^{11}-\frac{N^1 N^1}{N^2}=\frac{(a+b)^2}{a^2-1}<0.
\end{align*}

Using the formula for the determinant of block matrices, we obtain the
expression for the determinant of metric (\ref{eadmme}):
\begin{equation}                                        \label{edeadm}
  \det g_{\al\bt}=N^2\det g_{\mu\nu}.
\end{equation}
We hence have the expression for the volume element
\begin{equation}                                        \label{evolel}
  e=N\hat e,~~~~e=\sqrt{|\det g_{\al\bt}|},~~~~\hat e=\sqrt{|\det g_{\mu\nu}|}.
\end{equation}
This formula is a generalization of the well-known school rule:
the volume of a prism is equal to the product of the base area and the
height. In this case, the base area is $\hat e$ and the height is
the lapse function $N$.

The following formulas, which can be varified straightforwardly, are useful
for calculations:
\begin{equation}                                        \label{eadmef}
\begin{split}
  g^{00}g^{\mu\nu}-g^{0\mu}g^{0\nu}&=\frac{\hat g^{\mu\nu}}{N^2},
\\
  g^{\s\mu}g^{0\nu}-g^{\s\nu}g^{0\mu}&
  =\frac{N^\mu\hat g^{\s\nu}-N^\nu\hat g^{\s\mu}}{N^2},
\\
  g^{\mu\nu}g_{\nu\s}&=\dl^\mu_\s+\frac{N^\mu N_\s}{N^2},
\\
  g^{\mu\nu}g_{\mu\nu}&=n-1+\frac{N^\mu N_\mu}{N^2}.
\end{split}
\end{equation}
\section{Geometry of hypersurfaces                    \label{sgeohy}}
In the Hamiltonian formulation of gravity models, we take the space-time
as a family of spacelike hypersurfaces $x^0=\const$ parameterized by
time. In other words, in each instant, the space is a hypersurface
embedded in the space-time. It is useful to know what geometry arises on
spacelike hypersurfaces because equations of gravity models define geometry
of the whole space-time. In the present section, we approach this
problem from a general standpoint, assuming that an arbitrary affine
geometry is given on the embedding manifold without assuming that
the metric signature is Lorentzian.

We consider an $(n-1)$-dimensional hypersurface $\MU$ embedded in a
$n$-dimensional manifold $\MM$:
\begin{equation}                                          \label{eimmum}
  f:~~~~\MU\rightarrow \MM.
\end{equation}
We let $x^\al$, $\al=0,1,\dotsc,n-1$, and $u^i$, $i=1,\dotsc,n-1$, denote
the respective coordinates on $\MM$ and $\MU$. Then the embedding of $\MU$ in
$\MM$ is locally given by $n$ functions $x^\al(u)$, which we assume to be
sufficiently smooth. An arbitrary vector field $\lbrace X^i\rbrace\in\MT(\MU)$
on the hypersurface is mapped on the vector field
$\lbrace X^\al\rbrace\in\MT(\MM)$ on $\MM$ by the map differential
\begin{equation*}
  f_*:~~X=X^i\pl_i\in\MT(\MU)~~\rightarrow~~Y=Y^\al\pl_\al\in\MT(\MM),
\end{equation*}
where
\begin{equation*}
  Y^\al=e^\al{}_i X^i,~~~~e^\al{}_i=\pl_i x^\al.
\end{equation*}
The Jacobi matrix $e^\al{}_i$ of the transformation $f$ is rectangular of the
size $n\times(n-1)$, has the rank $n-1$, and is obviously irreversible.
It is defined not on the whole manifold $\MM$ but only on the hypersurface
$\MU$. In addition, we note that the Jacobi matrix is a vector and covector
with respect to the respective coordinates transformations on $\MM$ and $\MU$.
The pullback of the map $f$ maps each covector field on the image
$f(\MM)\subset\MM$ in the covector field on $\MU$:
\begin{equation*}
  f^*:~~A=dx^\al A_\al\in\MT^*(\MM)~~\rightarrow~~B=du^i B_i\in\MT^*(\MU),
\end{equation*}
where
\begin{equation*}
  B_i=A_\al e^\al{}_i.
\end{equation*}
We identify the hypersurface $\MU$ with its image $\MU=f(\MU)\subset\MM$
in what follows.

The 1-form $n=dx^\al n_\al$ defined on the hypersurface $\MU$ by the system of
algebraic equations
\begin{equation}                                         \label{eortre}
  n_\al e^\al{}_i=0,~~~~i=1,\dotsc,n-1,
\end{equation}
specifies the field of $(n-1)$-dimensional subspaces tangent to $\MU$ in
the tangent bundle $\MT(\MM)$. These equations have a unique solution up to
multiplication on an arbitrary nonzero function because the rank of the
Jacobi matrix is equal to $n-1$ as a consequence of the definition of
embedding.

The Jacobi matrix $e^\al{}_i$ defines a set of $n-1$ vectors $e_i=e^\al{}_i\pl_\al$
in the tangent spaces $\MT_x(\MM)$, $x\in\MU$; these vectors form the
basis of the space tangent to the hypersurface.

This is all we can say about a hypersurface $\MU$ if there is only embedding
(\ref{eimmum}). The theory becomes much richer in content if there are
additional structures on $\MM$. We discuss this question in detail.

Let the affine geometry be given on $\MM$, i.e., a metric $g_{\al\bt}$ and
an affine connection $\G_{\al\bt}{}^\g$. We consider what geometry arises
on the hypersurface $\MU\subset\MM$. The pullback of the map $f^*$ induces
a unique metric on the hypersurface:
\begin{equation}                                        \label{eindme}
  f^*:~~g_{\al\bt}~~\rightarrow~~g_{ij}=g_{\al\bt}e^\al{}_i e^\bt{}_j.
\end{equation}
The existence of the respective metrics $g_{\al\bt}$ and $g_{ij}$ on $\MM$
and $\MU$ allows lowering and raising the indices of the Jacobi matrix:
\begin{equation*}
  e_\al{}^i=g_{\al\bt}e^\bt{}_j g^{ij},
\end{equation*}
where $g^{ij}$ is the metric inverse to $g_{ij}$.
This matrix projects an arbitrary vector from $\MT_x(\MM)$, $x\in\MU$, into
the space tangent to the hypersurface $\MT(\MU)$
\begin{equation*}
  X^\al~~\rightarrow~~X^i=X^\al e_\al{}^i.
\end{equation*}

We now define the connection on the hypersurface $\MU\subset\MM$ by the relation
\begin{equation}                                        \label{edefco}
  \hat\nb_i X^k=(\nb_\al X^\bt)e^\al{}_i e_\bt{}^k.
\end{equation}
Opening this relation leads to the expression for the induced connection
on the hypersurface $\MU$ in the coordinate form:
\begin{equation}                                        \label{ecocou}
  \hat\G_{ij}{}^k=(\pl^2_{ij} x^\g+\G_{\al\bt}{}^\g e^\al{}_i e^\bt{}_j)e_\g{}^k.
\end{equation}
This connection is unique. We note that if the original connection $\G_{\al\bt}{}^\g$
is symmetric, then the induced connection $\hat\G_{ij}{}^k$ is also symmetric.
As a consequence of Eq.\ (\ref{ecocou}), the torsion tensor
$T_{\al\bt}{}^\g=\G_{\al\bt}{}^\g-\G_{\bt\al}{}^\g$ on $\MM$ induces the torsion
on the hypersurface
\begin{equation}                                          \label{eindto}
  T_{ij}{}^k=T_{\al\bt}{}^\g e^\al{}_i e^\bt{}_j e_\g{}^k.
\end{equation}
Furthermore, the connection on $\MU$ is uniquely defined only in the case where
the metric is given on $\MM$ in addition to the connection. All geometric objects
related to the hypersurface and constructed using only the induced metric
$g_{ij}$ and connection $\hat\G_{ij}{}^k$ are marked by the hat in what follows.

The metric $g_{\al\bt}$ and connection $\G_{\al\bt}{}^\g$ on $\MM$ define the
unique metric $g_{ij}$ and connection $\hat\G_{ij}{}^k$ on the hypersurface
$\MU\subset\MM$. The converse statement is not true. If a metric and
a connection are given on the hypersurface $\MU$, then they do not induce the
geometry on $\MM$ uniquely. This is clear because the dimension of the hypersurface
is less than that of the manyfold itself.

Straightforward calculations yield the expression for the covariant
derivative of the induced metric on the hypersurface:
\begin{equation*}
  \hat\nb_i g_{jk}=\pl_i g_{ij}-\hat\G_{ij}{}^l g_{lk}-\hat\G_{ik}{}^l g_{jl}
  =(\nb_\al g_{\bt\g})e^\al{}_i e^\bt{}_j e^\g{}_k.
\end{equation*}
This relation gives the expression for the nonmetricity tensor on the hypersurface
\begin{equation*}
  Q_{ijk}=Q_{\al\bt\g}e^\al{}_i e^\bt{}_j e^\g{}_k.
\end{equation*}
In particular, if the connection $\G_{\al\bt}{}^\g$ on $\MM$ is metrical
$(Q_{\al\bt\g}=0)$, then the induced connection $\hat\G_{ij}{}^k$ on $\MU$
is also metrical $(Q_{ijk}=0)$.

The existence of metric $g_{\al\bt}$ allows forming the unit vector field
$n=n^\al\pl_\al$ orthogonal to the hypersurface. It was already noted that the
system of equations $n_\al e^\al{}_i=0$ defines the 1-form $dx^\al n_\al$ up
to multiplication on an arbitrary scalar function. We use this arbitrariness
for the vector $n^\al=g^{\al\bt}n_\bt$ to have the unit length in every point:
$(n,n)=n^\al n^\bt g_{\al\bt}=1$. This vector is orthogonal to all vectors
tangent to the hypersurface by construction:
\begin{equation}                                           \label{eortnv}
  (n,e_i)=n^\al e^\bt{}_i g_{\al\bt}=n_\al e^\al{}_i=0.
\end{equation}

If the hypersurface is given on a manifold, then there is a natural basis
$\lbrace n,e_i\rbrace$ in the tangent space $\MT(\MM)$ defined by this
hypersurface. This basis is defined only in points of the hypersurface but
not on the whole manifold. The dual basis
$\lbrace n=dx^\al n_\al,e^i=dx^\al e_\al{}^i\rbrace$ in the cotangent space
$\MT^*(\MM)$ corresponds to it. Then an arbitrary vector $X$ and a 1-form $A$
can be decomposed with respect to this basis:
\begin{equation*}
\begin{split}
  X^\al&=X^\bot n^\al+X^i e^\al{}_i,~~~~X^\bot=X^\al n_\al,~~X^i=X^\al e_\al{}^i,
\\
  A_\al&=A_\bot n_\al+A_i e_\al{}^i,~~~~~A_\bot=A_\al n^\al,~~~A_i=A_\al e^\al{}_i.
\end{split}
\end{equation*}
A tensor of an arbitrary rank can be decomposed similarly. In particular,
a covariant second rank tensor has the decomposition
\begin{equation*}
  A_{\al\bt}=A_{\bot\bot}n_\al n_\bt +A_{\bot i}n_\al e^i{}_\bt
  +A_{i\bot}e^i{}_\al n_\bt+A_{ij}e^i{}_\al e^j{}_\bt,
\end{equation*}
where
\begin{equation*}
  A_{\bot\bot}=A_{\al\bt}n^\al n^\bt,~~A_{\bot i}=A_{\al\bt}n^\al e^\bt{}_i,~~
  A_{i\bot}=A_{\al\bt}e^\al{}_i n^\bt,~~A_{ij}=A_{\al\bt}e^\al{}_i e^\bt{}_j.
\end{equation*}

We can easily verify that the decomposition of the metric is essentially simpler:
\begin{equation}                                        \label{emetde}
  g_{\al\bt}=n_\al n_\bt+e_\al{}^i e_\bt{}^j g_{ij}.
\end{equation}
A similar decomposition holds for the inverse metric:
\begin{equation}                                        \label{einmer}
  g^{\al\bt}=n^\al n^\bt+e^\al{}_i e^\bt{}_j g^{ij}.
\end{equation}
The summation over Latin indices for the Jacobi matrix follows from the
definition of the inverse metric $g^{\al\bt}g_{\bt\g}=\dl^\al_\g$:
\begin{equation}                                        \label{esulae}
  e^\al{}_i e_\bt{}^i=\dl^\al_\bt-n^\al n_\bt.
\end{equation}

As a consequence of Eq.\ (\ref{eortnv}) and the definition of the inverse
induced metric $g^{ij}g_{jk}=\dl^i_k$, we have the equality
\begin{equation}                                        \label{esugre}
  e^\al{}_i e_\al{}^j=\dl_i^j,
\end{equation}
where the summation is over the Greek indices. Using this rule
we obtain the representation for the inverse induced metric
\begin{equation*}
  g^{ij}=g^{\al\bt}e_\al{}^i e_\bt{}^j
\end{equation*}
which follows from (\ref{einmer}). Metric (\ref{emetde}) and its inverse
(\ref{einmer}) in the basis $n,e_i$ have the block diagonal form:
\begin{equation*}
\begin{pmatrix}  1 & 0 \\ 0 & g_{ij} \end{pmatrix},~~~~
\begin{pmatrix}  1 & 0 \\ 0 & g^{ij} \end{pmatrix}.
\end{equation*}
This allows raising and lowering the corresponding indices. For example, if
$X_\al=X^\bt g_{\bt\al}$, then $X_\bot=X^\bot$ and $X_i=X^j g_{ji}$.

Induced metric (\ref{eindme}) and connection (\ref{ecocou}) define the
internal geometry of the hypersurface $\MU\subset\MM$. In particular, the
induced connection yields the internal curvature tensor of the hypersurface:
\begin{equation*}
  \hat R_{ijk}{}^l(\hat\G)=\pl_i\hat\G_{jk}{}^l
  -\hat\G_{ik}{}^m\hat\G_{jm}{}^l-(i\leftrightarrow j).
\end{equation*}

The embedding $f$ of the hypersurface allows defining an additional
important object which is called the external curvature of the hypersurface,
\begin{equation}                                        \label{excuhy}
  K_{ij}=-\nb_\al n_\bt e^\al{}_i e^\bt{}_j,
\end{equation}
which is equal to the covariant derivative of the normal projected on the
space tangent to the hypersurface up to a sign. In contrast to the internal
curvature tensor, the external curvature is a second-rank tensor, which has
no symmetry in the indices in the general case. This tensor characterizes the
variation of the normal when it is translated parallel along a curve on the
hypersurface. Expanding this definition and using (\ref{eortre}) we obtain
\begin{equation*}
  K_{ij}=n_\al(\pl^2_{ij} x^\al+\G_{\bt\g}{}^\al e^\bt{}_i e^\g{}_j).
\end{equation*}
The antisymmetric part of the external curvature tensor is given by
the torsion tensor,
\begin{equation}                                      \label{etonoc}
  K_{ij}-K_{ji}=2K_{[ij]}
  =n_\al T_{\bt\g}{}^\al e^\bt{}_i e^\g{}_j=T_{ij}{}^\bot.
\end{equation}
As a consequence, the external curvature is symmetrical if and only if the
connection $\G_{\bt\g}{}^\al$ has no torsion.

The covariant derivative of the Jacobi matrix is
\begin{equation}                                      \label{ecojam}
  \nb_i e^\al{}_j=e^\bt{}_i(\pl_\bt e^\al{}_j+\G_{\bt\g}{}^\al e^\g{}_j)
  -\hat\G_{ij}{}^k e^\al{}_k,
\end{equation}
where we use the connection $\G_{\al\bt}{}^\g$ on the whole manifold along
with the connection $\hat\G_{ij}{}^k$ on the hypersurface. Simple calculations
show that this covariant derivative has only the normal component and is
proportional to the external curvature:
\begin{equation}                                      \label{enexcu}
  \nb_i e^\al{}_j=n^\al K_{ij}.
\end{equation}
This relation is known as the Gauss--Weingarten formula. As a consequence,
we have one more representation for the external curvature tensor:
\begin{equation}                                        \label{excuco}
  K_{ij}=n_\al\nb_i e^\al{}_j.
\end{equation}

The full curvature tensor $R_{\al\bt\g\dl}$ of a manifold $\MM$ projected
on the hypersurface can be expressed in terms of the internal curvature tensor
$\hat R_{ijkl}$ constructed for induced metric (\ref{eindme}) and connection
(\ref{ecocou}) and the external curvature tensor. For this, we consider the
commutator of covariant derivatives of a vector field, which is given by
the curvature and torsion tensors,
\begin{equation}                                       \label{ecoman}
\begin{split}
  [\nb_\al,\nb_\bt]X_\g&=-R_{\al\bt\g}{}^\dl X_\dl-T_{\al\bt}{}^\dl\nb_\dl X_\dl=
\\
  &=-R_{\al\bt\g l}X^l-R_{\al\bt\g\bot}X^\bot-T_{\al\bt}{}^l\nb_l X_\g
  -T_{\al\bt}{}^\bot\nb_\bot X_\g,
\end{split}
\end{equation}
where we first compute the covariant derivatives in the right-hand side and
then project them on the hypersurface:
$\nb_l X_\g=e^\al{}_l\nb_\al X_\g$, $\nb_\bot X_\g=n^\al\nb_\al X_\g$.
To project this relation on the hypersurface, we first project the covariant
derivative:
\begin{equation*}
  \nb_i X_j=e^\al{}_i(\nb_\al X_\bt)e^\bt{}_j
  =\hat\nb_i X_j-X_\bot K_{ij},
\end{equation*}
where
\begin{equation*}
  \hat\nb_i X_j=\pl_i X_j-\hat\G_{ij}{}^k X_k
\end{equation*}
is $(n-1)$-dimensional covariant derivative on the hypersurface.
The second covariant derivative is projected similarly:
\begin{equation*}
\begin{split}
  \nb_i\nb_j X_k&=\hat\nb_i\nb_j X_k-\nb_\bot X_k K_{ij}-\nb_j X_\bot K_{ik}=
\\
  &=\hat\nb_i\hat\nb_j X_k-\hat\nb_iX_\bot K_{jk}-\hat\nb_jX_\bot K_{ik}
  -X_\bot\hat\nb_i K_{jk}-\nb_\bot X_k K_{ij},
\end{split}
\end{equation*}
where
\begin{equation*}
  \nb_i K_\bot=e^\al{}_i(\nb_\al X_\bt)n^\bt
  =\hat\nb_i X_\bt+X^jK_{ji}.
\end{equation*}
The antisymmetrization of the obtained expression in the indices $i,j$
yields the projection of commutator (\ref{ecoman}) on the hypersurface:
\begin{equation*}
  [\nb_i,\nb_j]X_k=-R_{ijkl}X^l-R_{ijk\bot}X^\bot
  -T_{ij}{}^l\nb_lX_k-T_{ij}{}^\bot\nb_\bot X_k.
\end{equation*}
Taking the independence of $X^l$ and $X^\bot$ and expressions (\ref{eindto})
and (\ref{etonoc}) for the torsion and curvature tensors into account, we
obtain the expressions for the projections of the full curvature tensor on
the hypersurface:
\begin{align}                                           \label{ecuhyc}
  R_{ijkl}&=\hat R_{ijkl}+K_{ik}K_{jl}-K_{jk}K_{il},
\\                                                      \label{ecunhy}
  R_{ijk\bot}&=\hat\nb_i K_{jk}-\hat\nb_j K_{ik}+T_{ij}{}^lK_{lk}.
\end{align}
The obtained relations are generalizations of the Gauss--Peterson--Codazzi
equations to the case where an arbitrary affine geometry with nonzero torsion
and nonmetricity is given on the embedding manifold $\MM$ instead of a
Riemannian geometry.

To conclude this section, we compute the normal components $G_{\bot\bot}$ and
$G_{\bot i}$ of the Einstein tensor
\begin{equation*}
  G_{\al\bt}=R_{\al\bt}-\frac12g_{\al\bt}R.
\end{equation*}
in the Riemannian geometry where the torsion and nonmetricity equal zero.
First, we compute the scalar curvature
\begin{equation*}
  R=g^{\al\g}g^{\bt\dl}R_{\al\bt\g\dl}=2R_{\bot\bot}+g^{ik}g^{jl}R_{ijkl},
\end{equation*}
where $R_{\bot\bot}=g^{ij}R_{i\bot j\bot}$ is the normal component of the
Ricci tensor and the expression for the inverse metric (\ref{einmer}) is
used. As a consequence of the Gauss--Peterson--Codazzi (\ref{ecuhyc}), we
obtain
\begin{equation*}
  g^{ik}g^{jl}R_{ijkl}=\hat R+K^2-K^{ij}K_{ij},
\end{equation*}
where $\hat R$ is the scalar internal curvature of the hypersurface and
$K=g^{ij}K_{ij}$ is the scalar external curvature of the hypersurface.
As a consequence, we obtain the expressions for the normal components
of the Einstein tensor
\begin{equation}                                        \label{einzeq}
\begin{split}
  G_{\bot\bot}&=-\frac12(\hat R+K^2-K^{ij}K_{ij}),
\\
  G_{\bot i}&=\hat\nb_jK_i{}^j-\nb_iK.
\end{split}
\end{equation}
It is important that these components of the Einstein tensor do not contain
derivatives normal to the hypersurface $\nb_\bot$ of the induced metric and
external curvature tensor at all. In the Hamiltonian language, this means
that time derivatives are absent and that Einstein's equations
$G_{\bot\bot}=0$ and $G_{\bot i}=0$ represent the constraints because
the components of the external curvature $K^{ij}$ are shown to be
proportional to the momenta canonically conjugate to the induced metric
$g_{ij}$ in the next section.
\section{Curvature in the ADM-parameterization of the metric}
The ADM parameterization of metric (\ref{eadmme}) is convenient for
the canonical formulation of general relativity in which the metric
components $g_{\al\bt}$ and canonically conjugate momenta $p^{\al\bt}$
are the independent variables. The passage from the Lagrangian to the
Hamiltonian needs a relatively tedious calculations, which are given here.

To essentially simplify calculations, we use the results of the preceding
section. Namely, the sections $x^0=\const$ of the space-time $\MM$ yield
the family of the hypersurfaces $\MU\subset\MM$ which are spacelike by the
assumption. We take the space coordinates as the coordinates on the
hypersurfaces
\begin{equation*}
  \lbrace u^i\rbrace \rightarrow\lbrace x^\mu\rbrace.
\end{equation*}
As a consequence, we loose the freedom of independent coordinate
transformations on the space-time $\MM$ and on the spacelike hypersurface
$\MU$, but many formulas became simpler.

The Jacobi matrix of the hypersurface embedding in the case under
consideration is
\begin{equation*}
  \lbrace e^\al{}_i\rbrace\rightarrow\lbrace 0_\nu,\dl^\mu_\nu\rbrace,~~~~
  \lbrace e_\al{}^i\rbrace\rightarrow\lbrace N^\mu,\dl^\mu_\nu\rbrace,
\end{equation*}
where $0_\nu$ denotes the row consisting of $n-1$ zeroes. The embedding
induces the metric $g_{\mu\nu}$ on the hypersurfaces according to
formula (\ref{eindme}).

We now construct the vector field $n=n^\al\pl_\al$ orthogonal to the family
of hypersurfaces $x^0=\const$. As a consequence of orthogonality condition
$(n,X)=0$, where $X=X^\mu\pl_\mu$ is a vector tangent to the section $x^0=0$,
we obtain
\begin{equation*}
  n=n^0(\pl_0-N^\mu\pl_\mu).
\end{equation*}
Moreover, if we set $n^0=1/N$, then the length of the normal vector is equal to
unity $(n^2=1)$. The unit vector orthogonal to a section $x^0=\const$ has the form
\begin{equation}                                        \label{enovei}
  n=\frac1N(\pl_0-N^\mu\pl_\mu)
\end{equation}
and is always timelike. The corresponding orthonormal 1-form is
\begin{equation}                                        \label{eoronf}
  n=dx^0N.
\end{equation}

An arbitrary vector on $\MM$ can be decomposed with respect to the
basis $\lbrace n,e_\mu\rbrace$. In particular, we have the decompositions
for vectors and 1-forms
\begin{equation*}
  X^\al=X^\bot n^\al+\tilde X^\mu e_\mu{}^\al,~~~~
  X_\al=X_\bot n_\al+\tilde X_\mu e^\mu{}_\al,
\end{equation*}
where
\begin{align*}
  X^\bot&=X^0N, & \tilde X^\mu&=X^0 N^\mu+X^\mu,
\\
  X_\bot&=\frac1N(X_0-N^\mu X_\mu), & \tilde X_\mu&=X_\mu.
\end{align*}

The representations for the metric (\ref{emetde}) of the whole space-time
and its inverse (\ref{einmer}) are
\begin{equation}                                        \label{emettr}
\begin{split}
  g_{\al\bt}&=n_\al n_\bt+g_{\mu\nu}e_\al{}^\mu e_\bt{}^\nu,
\\
  g^{\al\bt}&=n^\al n^\bt+\hat g^{\mu\nu}e_\mu{}^\al e_\nu{}^\bt.
\end{split}
\end{equation}

Connection (\ref{ecocou}) induced on the hypersurfaces is the Christoffel
symbols $\hat\G_{\mu\nu}{}^\rho$ computed for the space metric $g_{\mu\nu}$.

In the ADM-parameterization of the metric, external curvature tensor
(\ref{excuhy}) on a hypersurface $x^0=\const$ has the form
\begin{equation}                                        \label{excurv}
  K_{\mu\nu}=\G_{\mu\nu}{}^0N=\frac1{2N}(\hat\nb_\mu N_\nu
  +\hat\nb_\nu N_\mu-\dot g_{\mu\nu}),
\end{equation}
where the dot denotes the differentiation with respect to time,
$\dot g_{\mu\nu}=\pl_0 g_{\mu\nu}$ and
$\hat\nb_\mu N_\nu=\pl_\mu N_\nu-\hat\G_{\mu\nu}{}^\rho N_\rho$.
The external curvature tensor is symmetrical ($K_{\mu\nu}=K_{\nu\mu}$) because
the torsion vanishes in the metric formulation of general relativity.
In what follows, we need the trace of the external curvature tensor
\begin{equation*}
  K=K_\mu{}^\mu=\hat g^{\mu\nu}K_{\mu\nu}.
\end{equation*}
All time derivatives of the space part of the metric $\dot g_{\mu\nu}$ are
conveniently expressed in terms of $K_{\mu\nu}$ when computing the curvature
tensor $R_{\al\bt\g\dl}$ of the space-time $\MM$. Moreover, to exclude the second
time derivatives $\ddot g_{\mu\nu}$, we need the time derivative of the external
curvature tensor
\begin{equation*}
  \dot K_{\mu\nu}=\frac1{2N}\left[\hat\nb_\mu\dot N_\nu
  +\hat\nb_\nu\dot N_\mu-\ddot g_{\mu\nu}-N^\rho(\hat\nb_\mu\dot g_{\nu\rho}
  +\hat\nb_\nu\dot g_{\mu\rho}-\hat\nb_\rho\dot g_{\mu\nu})\right]
  -\frac{\dot N}NK_{\mu\nu},
\end{equation*}
where
\begin{equation*}
  \hat\nb_\mu\dot N_\nu=\pl_\mu\dot N_\nu-\hat\G_{\mu\nu}{}^\rho\dot N_\rho
\end{equation*}
and where we must substitute the expression for $\dot g_{\mu\nu}$ in terms of
$K_{\mu\nu}$.

We now compute the curvature tensor $R_{\al\bt\g\dl}$ for the metric of
form (\ref{eadmme}). Straightforward calculations yield the linearly
independent Christoffel symbols
\begin{equation}                                        \label{eadmch}
\begin{split}
  \G_{00}{}^0&=\frac1N\left(\dot N+N^\rho\pl_\rho N
  +N^\rho N^\s K_{\rho\s}\right),
\\
  \G_{00}{}^\mu&=\hat g^{\mu\nu}(\dot N_\nu-N\pl_\nu N
  -N^\rho\hat\nb_\nu N_\rho)
  -\frac{N^\mu}N\left(\dot N+N^\rho\pl_\rho N+N^\rho N^\s K_{\rho\s}\right),
\\
  \G_{0\mu}{}^0&=\frac1N\left(\pl_\mu N+N^\nu K_{\mu\nu}\right),
\\
  \G_{0\mu}{}^\nu&=\hat\nb_\mu N^\nu-NK_\mu{}^\nu-\frac{N^\nu}N
  (\pl_\mu N+N^\rho K_{\mu\rho}),
\\
  \G_{\mu\nu}{}^0&=\frac1NK_{\mu\nu},
\\
  \G_{\mu\nu}{}^\rho&=\hat\G_{\mu\nu}{}^\rho
  -\frac{N^\rho}N K_{\mu\nu}.
\end{split}
\end{equation}
In what follows, we need the following combinations of the Christoffel symbols
\begin{equation*}
  \G_\al=\G_{\al\bt}{}^\bt,~~~~g^{\bt\g}\G_{\bt\g}{}^\al.
\end{equation*}
Simple calculations give
\begin{equation}                                        \label{eadmlc}
\begin{split}
  \G_0&=\frac{\dot N}N+\hat\nb_\mu N^\mu-NK,
\\
  \G_\mu&=\hat\G_\mu+\frac{\pl_\mu N}N,
\\
  g^{\bt\g}\G_{\bt\g}{}^0&=\frac1NK
  +\frac1{N^3}(\dot N-N^\mu\pl_\mu N),
\\
  g^{\bt\g}\G_{\bt\g}{}^\mu&=\left(\hat g^{\rho\s}
  +\frac{N^\rho N^\s}{N^2}\right)\hat\G_{\rho\s}{}^\mu-\frac{N^\mu}N K
  -\frac{N^\mu}{N^3}(\dot N-N^\rho\pl_\rho N)+
\\
  &+\frac1{N^2}\hat g^{\mu\rho}(\dot N_\rho-N\pl_\rho N
  -N^\s\hat\nb_\rho N_\s-2N^\s\hat\nb_\s N_\rho+2NN^\s K_{\rho\s}).
\end{split}
\end{equation}
We also write the formulas for the time derivatives of the Christoffel symbols
\begin{equation}                                        \label{echtid}
\begin{split}
  \pl_0\hat\G_{\mu\nu\rho}&=
  \frac12(\hat\nb_\mu\dot g_{\nu\rho}+\hat\nb_\nu\dot g_{\mu\rho}
  -\hat\nb_\rho\dot g_{\mu\nu})+\hat\G_{\mu\nu}{}^\s\dot g_{\rho\s},
\\
  \pl_0\hat\G_{\mu\nu}{}^\s&=
  \frac12\hat g^{\s\rho}(\hat\nb_\mu\dot g_{\nu\rho}
  +\hat\nb_\nu\dot g_{\mu\rho}-\hat\nb_\rho\dot g_{\mu\nu}),
\\
  \pl_0\hat\G_\mu&=\frac12\hat g^{\nu\rho}\hat\nb_\mu\dot g_{\nu\rho}.
\end{split}
\end{equation}
The time derivatives $\dot g_{\mu\nu}$ are also eliminated from these
expressions using relation (\ref{excurv}).

We now compute the linearly independent components of the curvature tensor:
\begin{equation}                                        \label{eadmcu}
\begin{split}
  R_{0\mu0\nu}=&-N\dot K_{\mu\nu}+\hat R_{\mu\rho\nu\s}N^\rho N^\s
  +NN^\rho(\hat\nb_\mu K_{\nu\rho}+\hat\nb_\nu K_{\mu\rho}
  -\hat\nb_\rho K_{\mu\nu})+
\\
  &+N\hat\nb_\mu\hat\nb_\nu N
  +K_{\mu\nu}N^\rho N^\s K_{\rho\s}+N(K_\mu{}^\rho\hat\nb_\nu N_\rho
  +K_\nu{}^\rho\hat\nb_\mu N_\rho)-
\\
  &-N^2K_\mu{}^\rho K_{\nu\rho}-N^\rho N^\s K_{\mu\rho}K_{\nu\s},
\\
  R_{\mu\nu\rho0}=&\hat R_{\mu\nu\rho\s}N^\s
  +N(\hat\nb_\mu K_{\nu\rho}-\hat\nb_\nu K_{\mu\rho})
  +(K_{\mu\rho}K_{\nu\s}-K_{\nu\rho}K_{\mu\s})N^\s,
\\
  R_{\mu\nu\rho\s}=&\hat R_{\mu\nu\rho\s}
  +K_{\mu\rho}K_{\nu\s}-K_{\mu\s}K_{\nu\rho},
\end{split}
\end{equation}
where we use the formula for the commutator of covariant derivatives
\begin{equation*}
  (\hat\nb_\mu\hat\nb_\nu-\hat\nb_\nu\hat\nb_\mu)N_\rho
  =-\hat R_{\mu\nu\rho\s}N^\s.
\end{equation*}

The components of the curvature tensor having at least one time index
seem simpler when expressed in the basis $n,e_\mu$:
\begin{equation}                                        \label{ecuorn}
\begin{split}
  R_{\bot\mu\bot\nu}&=\frac1N\left(-\dot K_{\mu\nu}+\hat\nb_\mu\hat\nb_\nu N
  +K_{\mu\rho}\hat\nb_\nu N^\rho+K_{\nu\rho}\hat\nb_\mu N^\rho
  -NK_{\mu\rho}K_\nu{}^\rho+N^\rho\hat\nb_\rho K_{\mu\nu}\right),
\\
  R_{\mu\nu\rho\bot}&=\hat\nb_\mu K_{\nu\rho}-\hat\nb_\nu K_{\mu\rho}.
\end{split}
\end{equation}
The components of the curvature tensor $R_{\mu\nu\rho\s}$ and $R_{\mu\nu\rho\bot}$
were actually obtained in the preceeding section (\ref{ecuhyc}),
(\ref{ecunhy}) without straightforward calculations.

The Ricci tensor has the linearly independent components
\begin{equation}                                        \label{eadmri}
\begin{split}
  R_{00}=&-N\hat g^{\mu\nu}\dot K_{\mu\nu}+\hat R_{\mu\nu}N^\mu N^\nu
  +NN^\mu(2\hat\nb_\nu K^\nu{}_\mu-\pl_\mu K)+N\hat\nb^\mu\hat\nb_\mu N+
\\
  &+N^\mu N^\nu K_{\mu\nu}K+2NK^{\mu\nu}\hat\nb_\mu N_\nu-N^2K^{\mu\nu}K_{\mu\nu}
  -2N^\mu N^\nu K_\mu{}^\rho K_{\nu\rho}+
\\
  &+\frac{N^\mu N^\nu}N\left(-\dot K_{\mu\nu}+N^\rho\hat\nb_\mu K_{\nu\rho}
  +\hat\nb_\mu\hat\nb_\nu N+2K_\mu{}^\rho\hat\nb_\nu N_\rho\right),
\\
  R_{0\mu}=&\frac{N^\nu}N\left(-\dot K_{\mu\nu}+N^\s\hat\nb_\nu K_{\mu\s}
  +\hat\nb_\mu\hat\nb_\nu N+K_\mu{}^\rho\hat\nb_\nu N_\rho
  +K_\nu{}^\rho\hat\nb_\mu N_\rho\right)
\\
  &+\hat R_{\mu\nu}N^\nu+N\left(\hat\nb_\nu K^\nu{}_\mu-\pl_\mu K\right)+
  +K_{\mu\nu}N^\nu K-2K_\mu{}^\rho K_{\nu\rho}N^\nu,
\\
  R_{\mu\nu}=&\hat R_{\mu\nu}+\frac1N\left(-\dot K_{\mu\nu}
  +\hat\nb_\mu\hat\nb_\nu N
  +K_\mu{}^\rho\hat\nb_\nu N_\rho+K_\nu{}^\rho\hat\nb_\mu N_\rho\right)+
\\
  &+\frac{N^\rho}N\hat\nb_\rho K_{\mu\nu}+K_{\mu\nu}K-2K_\mu{}^\rho K_{\nu\rho}.
\end{split}
\end{equation}
For reference, we also write the Ricci tensor components with respect to the
basis $n,e_\mu$:
\begin{equation}                                        \label{ericor}
\begin{split}
  R_{\bot\bot}=&-\frac1N\hat g^{\mu\nu}\dot K_{\mu\nu}
  +\frac1N\hat\nb^\mu\hat\nb_\mu N+\frac2N K^{\mu\nu}\hat\nb_\mu N_\nu
  -K^{\mu\nu}K_{\mu\nu}+\frac{N^\mu}N\pl_\mu K,
\\
  R_{\bot\mu}=&\hat\nb_\nu K^\nu{}_\mu-\pl_\mu K.
\end{split}
\end{equation}

Finally, we compute the scalar curvature:
\begin{equation}                                        \label{eadmsc}
\begin{split}
  R&=\hat R
  +\frac2N \left(-\hat g^{\mu\nu}\dot K_{\mu\nu}+\hat\nb^\mu\hat\nb_\mu N
  +2K^{\mu\nu}\hat\nb_\mu N_\nu+N^\mu\pl_\mu K\right)-3K^{\mu\nu}K_{\mu\nu}+K^2.
\end{split}
\end{equation}
\section{The Hamiltonian}
The scalar curvature contains second derivatives of the metric components
with respect to time as well as the space coordinates and is therefore not
suitable for canonically formulating general relativity. It suffices
to eliminate only the second time derivatives from the Lagrangian to obtain
a canonical formulation. The Lagrangian density takes the simplest form after
adding the boundary term:
\begin{equation}                                        \label{ehieil}
  {\cal L}_{\Sa\Sd\Sm}=N\hat eR
  +2\pl_0\left(\hat eK\right)
  -2\pl_\mu\left(\hat e\hat g^{\mu\nu}\pl_\nu N\right).
\end{equation}
Straightforward calculations yield the expression
\begin{equation}                                        \label{eadmla}
  {\cal L}_{\Sa\Sd\Sm}=N\hat e\left(K^{\mu\nu}K_{\mu\nu}-K^2+\hat R\right).
\end{equation}

The transition to the Hamiltonian formalism is now easy. First, the
ADM Lagrangian does not contain time derivatives of the lapse and shift
functions $N$ and $N_\mu$. This means that the theory contains $n$ primary
constraints
\begin{equation}                                        \label{eprcge}
  p^\bot=\frac{\pl {\cal L}_{\Sa\Sd\Sm}}{\pl\dot N}=0,~~~~
  p^\mu=\frac{\pl {\cal L}_{\Sa\Sd\Sm}}{\pl\dot N_\mu}=0,
\end{equation}
their number coinciding with the number of independent functions that
parametrize diffeomorphisms.

The momenta canonically conjugate to the space metric $g_{\mu\nu}$ are
proportional to the external curvature tensor,
\begin{equation}                                        \label{emomet}
  p^{\mu\nu}=\frac{\pl {\cal L}_{\Sa\Sd\Sm}}{\pl\dot g_{\mu\nu}}
  =-\frac1{2N}\frac{\pl {\cal L}_{\Sa\Sd\Sm}}{\pl\dot K_{\mu\nu}}
  =-\hat e\left(K^{\mu\nu}-\hat g^{\mu\nu}K\right).
\end{equation}
We note that the momenta are not tensors with respect to the coordinate
transformations $x^\mu$ but tensor densities of the degree $-1$, as is
the determinant of the vielbein, which degree is
\begin{equation*}
  \deg \hat e=-1
\end{equation*}
by definition.

To eliminate the velocities $\dot g_{\mu\nu}$ from the ADM Lagrangian, we decompose
the momenta on the irreducible components, extracting the trace from $p^{\mu\nu}$
\begin{equation}                                        \label{eirmom}
  p^{\mu\nu}=\tilde p^{\mu\nu}+\frac1{n-1}p\hat g^{\mu\nu},
\end{equation}
where we introduce the trace of the momenta
\begin{equation}                                        \label{exptrm}
  p=p^{\mu\nu}g_{\mu\nu}=\hat e(n-2)K
\end{equation}
and the symmetric traceless part
\begin{equation*}
  \tilde p^{\mu\nu}=\tilde p^{\nu\mu}
  =-\hat e\left(K^{\mu\nu}-\frac1{n-1}\hat g^{\mu\nu}K\right),
  ~~~~\tilde p^{\mu\nu}g_{\mu\nu}=0.
\end{equation*}
We can now solve Eq.\ (\ref{emomet}) for the velocities
using relation (\ref{excurv}),
\begin{equation*}
  \dot g_{\mu\nu}=\frac{2N}{\hat e}\left(p_{\mu\nu}
  -\frac1{(n-2)}pg_{\mu\nu}\right)+\hat\nb_\mu N_\nu+\hat\nb_\nu N_\mu.
\end{equation*}
Simple calculations yield the Hamiltonian density
\begin{equation}                                        \label{ehawid}
  \CH=p^{\mu\nu}\dot g_{\mu\nu}-\CL_{\Sa\Sd\Sm}
  =NH_\bot+N^\mu H_\mu+2\pl_\mu(p^{\mu\nu}N_\nu),
\end{equation}
where
\begin{equation}                                        \label{ecoadm}
\begin{split}
  H_\bot&=\frac1{\hat e}\left(p^{\mu\nu}p_{\mu\nu}-\frac1{(n-2)}p^2\right)
  -\hat e\hat R,
\\
  H_\mu&=-2\hat\nb_\nu p^\nu{}_\mu=-2\pl_\nu p^\nu{}_\mu
  +\pl_\mu g_{\nu\rho}p^{\nu\rho},
\end{split}
\end{equation}
and $p_{\mu\nu}=g_{\mu\rho}g_{\nu\s}p^{\rho\s}$. We note that the covariant
derivative of the momenta contains only one term with Christoffel symbols
because the momenta are tensor densities. The expressions for $H_\bot$ and
$H_\mu$ are proportional to the $G_{\bot\bot}$ and $G_{\bot\mu}$ components
of Einstein tensor (\ref{einzeq}), which justifies the chosen notations.

Dropping the divergence in the expression for the Hamiltonian density
(\ref{ehawid}), we obtain the final expression for the Hamiltonian:
\begin{equation}                                        \label{eadmha}
  H_{\Sa\Sd\Sm}=\int d\Bx\CH_{\Sa\Sd\Sm}=\int d\Bx(NH_\bot+N^\mu H_\mu).
\end{equation}

We now rewrite the expression for $H_\bot$ in terms of the irreducible
components of momenta:
\begin{equation*}
  H_\bot=\frac1{\hat e}\left[\tilde p^{\mu\nu}\tilde p_{\mu\nu}
  -\frac1{(n-1)(n-2)}p^2\right]-\hat e\hat R.
\end{equation*}
In particular, the quadratic part of the momenta in $H_\bot$ for $n\ge3$
is consequently not positive definite.
\section{Secondary constraints                        \label{salcog}}
To finish constructing the Hamiltonian formalism, we must analyze
the consistency of primary constraints (\ref{eprcge}) with the
equations of motion. The phase space of general relativity is described
by $n(n+1)$ conjugate coordinates and momenta: $(N,p^\bot)$, $(N_\mu,p^\mu)$,
$(g_{\mu\nu},p^{\mu\nu})$ on which the canonical equal-time Poisson bracket
\begin{equation}                                        \label{epoist}
  [N,p^{\prime\bot}]=\dl,~~~~[N_\mu,p^{\prime\nu}]=\dl_\mu^\nu\dl,~~~~
  [g_{\mu\nu},p^{\prime\rho\s}]=\dl_{\mu\nu}^{\rho\s}\dl,
\end{equation}
is given, where primed field variables are considered at a point
$\Bx'=(x^{\prime1},\dotsc,x^{\prime n-1})$. All fields are considered at
the same instant $t=x^0$. For brevity, we use the notation
\begin{equation}                                        \label{epoino}
\begin{split}
  \dl&=\dl^{(n-1)}(\Bx-\Bx')=\dl(x^1-x^{\prime1})\dotsc\dl(x^{n-1}-x^{\prime n-1}),
\\
  \dl&_{\mu\nu}^{\rho\s}=\frac12(\dl_\mu^\rho\dl_\nu^\s+\dl_\mu^\s\dl_\nu^\rho).
\end{split}
\end{equation}
for the $(n-1)$-dimensional $\dl$-function and the symmetric combination
of the Kronecker symbols in the right-hand sides of the Poisson brackets.
We write all $\dl$-functions on the right to distinguish them from a
field variation.

We now consider the Hamiltonian equations of motion for primary
constraints (\ref{eprcge}):
\begin{equation*}
  \dot p^\bot=[p^\bot,H_{\Sa\Sd\Sm}]=-H_\bot,~~~~
  \dot p^\mu=[p^\mu,H_{\Sa\Sd\Sm}]=-H^\mu.
\end{equation*}
The consistency of the primary constraints with the equations of motion
$\dot p^\bot=0$, $\dot p^\mu=0$ leads to the secondary constraints
\begin{equation}                                        \label{esecco}
  H_\bot=0,~~~~H_\mu=0,
\end{equation}
where $H_\bot=H^\bot$ and $H_\mu=g_{\mu\nu}H^\nu$. We note that the secondary
constraints are not tensors but tensor densities of degree $-1$. Moreover,
it is more convenient to consider the equivalent set of constraints with
lowered index $H_\mu$ instead of the constraints $H^\mu$. Below, we show
that these constraints define the generators of coordinate transformations
on the sections $x^0=\const$ and satisfy a simpler algebra.

The constraints $H_\mu$ are linear in the momenta and metric. The constraint
$H_\bot$ is quadratic in the momenta and nonpolynomial in the metric
$g_{\mu\nu}$ because it depends on the square root of the metric $\hat e$
and the inverse metric $\hat g^{\mu\nu}$. The last circumstance raises
essential technical difficulties in the perturbation theory.

The secondary constraints are independent on the canonical variables
$(N,p^\bot)$ and $(N_\mu,p^\mu)$, and they can be eliminated by considering the
$n(n-1)$-dimensional phase space of the variables $g_{\mu\nu}$ and $p^{\mu\nu}$
on which constraints (\ref{esecco}) are imposed. In this case, the lapse and
shift functions $N$ and $N_\mu$ are regarded as Lagrange multipliers
in the problem for the conditional extremum for the action
\begin{equation*}
  S=\int d^n x(p^{\mu\nu}\dot g_{\mu\nu}-\CH_{\Sa\Sd\Sm}).
\end{equation*}

Because Hamiltonian (\ref{eadmha}) in general relativity is equal to
a linear combination of the secondary constraints, we must compute the
Poisson brackets of the constraints between themselves to analyze the
consistency of secondary constraints (\ref{esecco}) with the equations of
motion. The constraint algebra in general relativity is well known,
\begin{align}                                           \label{econoo}
  [H_\bot,H'_\bot]&=-(H_\mu\hat g^{\mu\nu}+H'_\mu\hat g^{\prime\mu\nu})\dl_\nu,
\\                                                      \label{econom}
  [H_\bot,H'_\mu]&=-H'_\bot\dl_\mu,
\\                                                      \label{econmn}
  [H_\mu,H'_\nu]&=-H_\nu\dl_\mu-H'_\mu\dl_\nu,
\end{align}
where we use the shorthand notation for the derivative of the $\dl$-function,
\begin{equation*}
  \dl_\mu=\frac\pl{\pl x^{\prime\mu}}\dl(x'-x).
\end{equation*}

Straightforward calculations of constraint algebra (\ref{econoo})--(\ref{econmn})
are very tedious. This algebra was first written by Dirac \cite{Dirac51} using
symmetry considerations. The assumptions on the form of the constraints made in
the course of derivation are not satisfied in general relativity, and the
existence of the corresponding canonical transformation is now questionable.
Therefore, Dirac's derivation of the constraints algebra cannot be considered
satisfactory.

Two Poisson brackets (\ref{econom}) and (\ref{econmn}) can in fact be found
without straightforward calculations. For this, we consider the functional
\begin{equation*}
  T_u=-\int d\Bx~u^\mu H_\mu,
\end{equation*}
where $u^\mu(x)$ is an infinitesimal vector field. Calculating the Poisson
brackets of the phase-space coordinates $g_{\mu\nu}$ and $p^{\mu\nu}$
with $T_u$ yields
\begin{equation*}
\begin{split}
  \dl_u g_{\mu\nu}&=[g_{\mu\nu},T_u]=-\pl_\mu u^\rho g_{\rho\nu}
  -\pl_\nu u^\rho g_{\mu\rho}-u^\rho\pl_\rho g_{\mu\nu},
\\
  \dl_u p^{\mu\nu}&=[p^{\mu\nu},T_u]=\pl_\rho u^\mu p^{\rho\nu}
  +\pl_\rho u^\nu p^{\mu\rho}-\pl_\rho(u^\rho p^{\mu\nu}).
\end{split}
\end{equation*}
This means that the functional $T_u$, which is defined by the constraints $H_\mu$,
is the generator of general coordinate transformations on the hypersurfaces
$x^0=\const$. We recall that the momenta $p^{\mu\nu}$ are tensor densities of
degree $-1$. The algebra of general coordinate transformations is well known
and defined by the Poisson bracket (\ref{econmn}). We also can avoid computing
the Poisson bracket (\ref{econom}) explicitly. Its form follows from the fact
that the constraint $H_\bot$ is the scalar density of weight $-1$. Therefore,
only the Poisson bracket (\ref{econoo}) must be computed. These calculations,
being very cumbersome, were apparently first performed much later by DeWitt
\cite{DeWitt67A}. In the next section, we compute this Poisson bracket after
the canonical transformation, which casts the constraints into a polynomial
form,essentially simplifying the calculations.

We say that the constraints $H_\mu$ are kinematical because they define only
space diffeomorphisms. They are also independent of the coupling constants
in the action if there are any. The constraint $H_\bot$ is said to be dynamical
because it governs the evolution of the initial data in time and depends
essentially on the original action, in particular, on the coupling constants.

For comparison, we write the Poisson brackets of the constraints $H^\mu=0$
with a contravariant index which are equivalent to the constraints $H_\mu=0$:
\begin{equation*}
  [H^\mu,H^{\prime \nu}]=\left(\hat g^{\mu\nu}H^\rho
  +\hat g^{\prime\mu\nu}H^{\prime\rho}\right)\dl_\rho
  +\left(\hat g^{\mu\rho}\pl_\rho g_{\s\lm}\hat g^{\nu\s}
  -\hat g^{\nu\rho}\pl_\rho g_{\s\lm}\hat g^{\mu\s}\right)H^\lm\dl.
\end{equation*}
We see that this seems more complicated than bracket (\ref{econmn}).
\section{The canonical transformation                 \label{scatrg}}
The idea of the canonical transformation is as follows. The momenta $p^{\mu\nu}$
are reducible and decompose into the traceless part and the trace (\ref{eirmom}).
Usually, working with irreducible components is more convenient for
calculations because many terms automatically cancel. We pose the question:
``Is it possible to perform a canonical transformation such that the irreducible
components $\tilde p^{\mu\nu}$ and $p$ become new canonical momenta?''
This question is nontrivial because the decomposition of the momenta involves
the metric, whose components themselves are coordinates of the phase space.
The answer to this question is negative because the Poisson brackets between
the momenta are nonzero. For example, $[\tilde p^{\mu\nu},p']\ne0$. Nevertheless,
there is a canonical transformation such that the new momenta are proportional
to the irreducible components $\tilde p^{\mu\nu}$ and $p$. Constructing this
canonical transformation is our subject in this section.

We consider the canonical transformation
\begin{equation}                                               \label{ecatrn}
  (g_{\mu\nu},p^{\mu\nu})~~\rightarrow~~(k_{\mu\nu},P^{\mu\nu}), (\rho,P),
\end{equation}
to the new pairs of canonically conjugate coordinates and momenta with the
additional constraints on the coordinates $k_{\mu\nu}=k_{\nu\mu}$ and
conjugate momenta $P^{\mu\nu}=P^{\nu\mu}$
\begin{equation}                                                \label{econev}
  |\det k_{\mu\nu}|=1,~~~~P^{\mu\nu}k_{\mu\nu}=0,
\end{equation}
and $\rho>0$. We choose the space integral as the generator of the canonical
transformation
\begin{equation}                                        \label{egefun}
  F=-\int d\Bx~\rho^m k_{\mu\nu}p^{\mu\nu},~~~~m\in\MR,~~m\ne0,
\end{equation}
depending on the new coordinates $\rho,k_{\mu\nu}$, old momenta $p^{\mu\nu}$,
and the real parameter $m$. The old coordinates and new momenta are then
given by the variational derivatives (see, e.g., \cite{Goldst50})
\begin{align}                                           \label{eoldme}
  g_{\mu\nu}&=-\frac{\dl F}{\dl p^{\mu\nu}}=\rho^m k_{\mu\nu},
\\                                                      \label{eoldmo}
  P^{\mu\nu}&=-\frac{\dl F}{\dl k_{\mu\nu}}=\rho^m\tilde p^{\mu\nu},
\\                                                      \label{eoldmt}
  P&=-\frac{\dl F}{\dl\rho}=\frac m\rho p.
\end{align}
Computing the variational derivative with respect to $k_{\mu\nu}$, we use the
condition $|\det k_{\mu\nu}|=1$, which restricts the variations
$k^{\mu\nu}\dl k_{\mu\nu}=0$, where $k^{\mu\nu}$ is the tensor density inverse
to $k_{\mu\nu}$: $k^{\mu\nu}k_{\nu\s}=\dl^\mu_\s$. Hence, the vanishing of
momentum traces (\ref{econev}) follows automatically from the unit determinant
condition for the density $k_{\mu\nu}$ for generating functional (\ref{egefun}).
In Eq.\ (\ref{eoldmt}), we use relation (\ref{eoldme}).

In essence, the determinant of the metric raised to some power is singled
out from the metric,
\begin{equation*}
  \rho=|\det g_{\mu\nu}|^{\frac1{m(n-1)}},
\end{equation*}
as a consequence of (\ref{eoldme}). For brevity in what follows, we also call
the symmetrical tensor density with the unit determinant $k_{\mu\nu}$ the metric.

Variables (\ref{ecatrn}) for $n=4$ and $m=1/2$ were considered in \cite{Tate92},
although the canonical transformation was not noted there.

Straightforward calculations yield the expression for the scalar curvature
of the section $x^0=\const$ in the new coordinates:
\begin{equation}                                        \label{escpka}
  \hat R=\rho^{-m-2}\left[\rho^2R^{(k)}+m(n-2)\rho\pl_\mu(k^{\mu\nu}\pl_\nu\rho)
  +m(n-2)\left(m\frac{n-3}4-1\right)k^{\mu\nu}\pl_\mu\rho\pl_\nu\rho\right].
\end{equation}
The ``scalar curvature'' for the metric $k_{\mu\nu}$ takes a particular simple form,
\begin{equation}                                        \label{escuka}
  R^{(k)}=\pl^2_{\mu\nu}k^{\mu\nu}
  +\frac12k^{\mu\nu}\pl_\rho k_{\mu\s}\pl_\nu k^{\rho\s}
  -\frac14k^{\mu\nu}\pl_\mu k_{\rho\s}\pl_\nu k^{\rho\s}.
\end{equation}
This expression is not a scalar with respect to coordinate transformations of
$x^\mu$, because $k_{\mu\nu}$ is not a tensor but tensor density. But we note
that the group of diffeomorphisms of sections $x^0=\const$ has a subgroup
consisting of coordinates transformations of $x^\mu$ with a unit determinant.
The density $k_{\mu\nu}$ is a tensor and $R^{(k)}$ is a scalar with respect to
this subgroup.

As a consequence of the unit determinant of the metric, we obtain the
components of the inverse metric $k^{\mu\nu}$ as polynomials of degree
$n-2$ in the components $k_{\mu\nu}$,
\begin{equation*}
  k^{\mu\nu}=\frac1{(n-2)!}\hat\ve^{\mu\rho_1\dotsc\rho_{n-2}}
  \hat\ve^{\nu\s_1\dotsc\s_{n-2}}k_{\rho_1\s_1}\dotsc k_{\rho_{n-2}\s_{n-2}},
\end{equation*}
where $|\hat\ve^{\mu_1\dotsc\mu_{n-1}}|=1$ is the totally antisymmetric tensor
density of rank $n-1$. Therefore, the scalar curvature $R^{(k)}$ is polynomial
in the metric $k_{\mu\nu}$ as well as in its inverse $k^{\mu\nu}$.

The dynamical constraint in the new variables becomes
\begin{multline*}
  H_\bot=\rho^{-\frac{m(n-1)}2}\left[P^{\mu\nu}P_{\mu\nu}
  -\frac{\rho^2}{m^2(n-1)(n-2)}P^2\right]
\\
  -\rho^{\frac{m(n-1)}2-m-2}\left[\rho^2R^{(k)}
  +m(n-2)\rho\pl_\mu(k^{\mu\nu}\pl_\nu\rho)
  +m(n-2)\left(m\frac{n-3}4-1\right)k^{\mu\nu}\pl_\mu\rho\pl_\nu\rho\right],
\end{multline*}
where $P_{\mu\nu}=k_{\mu\rho}k_{\nu\s}P^{\rho\s}$.

We now analyze the possibility of choosing the constant $m$ such that the
dynamical constraint becomes polynomial. Both expression in square
brackets are polynomial in all dynamical variables. Because $n\ge3$, we need
the inequality $m<0$ to ensure a positive power of the density $\rho$ before
the first square bracket. In this case, the power of $\rho$ before the second
square bracket is negative. We therefore cannot ensure that the constraint
$H_\bot$ itself is polynomial by choosing the constant $m$. But a constraint
can be multiplied by an arbitrary nonzero factor without changing the
surface defined by the constraint in the phase space. The power of $\rho$
by which $H_\bot$ be multiplied is minimum when the powers of $\rho$ before
the square brackets are equal. We consequently obtain th equality
\begin{equation*}
  m=\frac2{n-2}.
\end{equation*}
Then multiplying the dynamical constraint
\begin{equation}                                        \label{etrakh}
  K_\bot=\rho^{\frac{n-1}{n-2}}H_\bot=\hat e H_\bot,
\end{equation}
we obtain the equivalent polynomial constraint
\begin{equation}                                        \label{enecon}
  K_\bot=P^{\mu\nu}P_{\mu\nu}-\frac{n-2}{4(n-1)}\rho^2P^2
  -\rho^2\check R=0,
\end{equation}
where the scalar curvature density
\begin{equation*}
  \check R(\rho,k)=R^{(k)}+2\frac{\pl_\mu(k^{\mu\nu}\pl_\nu\rho)}\rho
  -\frac{n-1}{n-2}\frac{k^{\mu\nu}\pl_\mu\rho\pl_\nu\rho}{\rho^2},~~~~
  \deg\check R=-\frac2{n-1},
\end{equation*}
is introduced. We note that the ``scalar'' curvature $R^{(k)}$ constructed for
the metric density $k_{\mu\nu}$ is not a scalar density. Therefore, using
$\check R$ instead of $R^{(k)}$ simplifies many formulas and calculations.

Multiplying the dynamical constraint $H_\bot$ by the nonzero factor leads to
modifying the Lagrange multiplier (the lapse function) by the inverse factor,
\begin{equation}                                         \label{elagtr}
  N\rightarrow\widetilde N=N\rho^{-\frac{n-1}{n-2}}.
\end{equation}
In turn, the modification of the Lagrange multipliers in general relativity is
equivalent to coordinate changes that do not change the physical content of
the theory.

The kinematical constraints in the new dynamical variables remain polynomial:
\begin{equation}                                        \label{ekicon}
  H_\mu=-2\nb_\nu(P^\nu{}_\mu)-\frac{n-2}{n-1}\nb_\mu(P\rho)=0,
\end{equation}
where $\nb_\mu$ is the covariant derivative constructed for the metric
\begin{equation*}
  g_{\mu\nu}=\rho^{\frac2{n-2}}k_{\mu\nu},
\end{equation*}
and indices are lowered using the tensor density
$P^\nu{}_\mu=P^{\nu\rho}k_{\rho\mu}$. It can be easily verified that
\begin{equation*}
  \nb_\mu\rho=0,~~~~\nb_\mu k_{\nu\rho}=0.
\end{equation*}
Lowering and raising indices with the metric density $k_{\mu\nu}$ therefore
commutes with the covariant differentiation operation.

We note that the covariant derivative of a tensor density $\phi$ of degree
$\deg\phi=r$ in our notation is given by the expression:
\begin{equation*}
  \nb_\mu\phi=\pl_\mu\phi+r\G_\mu\phi,~~~~
  \G_\mu=\G_{\nu\mu}{}^\nu=\frac{n-1}{n-2}\frac{\pl_\mu\rho}\rho.
\end{equation*}
All new canonical variables are tensor densities of the degrees
\begin{alignat*}{2}
  \deg k_{\mu\nu}&=\frac2{n-1}, &\qquad\deg\rho&=-\frac{n-2}{n-1},
\\
  \deg P^{\mu\nu}&=-\frac{n+1}{n-1},  &\deg P&=\frac1{n-1},
\end{alignat*}
and this should be taken into account for covariant differentiation.

We now compute the basic Poisson brackets for the new canonical variables,
which follow from explicit expressions (\ref{eoldme})--(\ref{eoldmt})
Only three brackets are nonzero:
\begin{align}                                           \label{eponer}
  [\rho,P']&=\dl,
\\                                                      \label{eponek}
  [k_{\mu\nu},P^{\prime\rho\s}]&=\left(\dl_{\mu\nu}^{\rho\s}
  -\frac1{n-1}k_{\mu\nu}k^{\rho\s}\right)\dl,
\\                                                      \label{elpopp}
  [P^{\mu\nu},P^{\prime\rho\s}]&=\frac1{n-1}(P^{\mu\nu}k^{\rho\s}
  -P^{\rho\s}k^{\mu\nu})\dl.
\end{align}
Poisson brackets (\ref{eponek}) and (\ref{elpopp}) do not have the canonical
form for the phase variables. This occures because the fields $k_{\mu\nu}$
and $P^{\mu\nu}$ are subjected to additional constraints (\ref{econev}).

Because the Hamiltonian
\begin{equation*}
  H=\int d\Bx(\widetilde NK_\bot+N^\mu H_\mu)
\end{equation*}
is polynomial in the new variables, the equations of motion are also polynomial.
Straightforward calculations with Poisson brackets (\ref{eponer})--(\ref{elpopp})
yield the equations of motion:
\begin{align*}
  \dot\rho=&-\frac{n-2}{2(n-1)}\widetilde N\rho^2P
  +\frac{n-2}{n-1}\rho\nb_\mu N^\mu,
\\
  \dot P=&\frac{n-2}{2(n-1)}\widetilde N\rho P^2+2\widetilde N\rho\check R
  +2\rho\Box\widetilde N
  +\frac1{n-1} P\nb_\mu N^\mu+N^\mu\nb_\mu P,
\\
  \dot k_{\mu\nu}=&2\widetilde NP_{\mu\nu}+\nb_\mu N_\nu
  +\nb_\nu N_\mu-\frac2{n-1} k_{\mu\nu}\nb_\rho N^\rho,
\\
  \dot P^{\mu\nu}=&-2\widetilde NP^{\mu\rho}P^\nu{}_\rho
  -\rho^2\left(\widetilde Nk^{\mu\rho}k^{\nu\s}\hat R_{\rho\s}
  -\frac1{n-1}\widetilde N\check Rk^{\mu\nu}
  +\nb^\mu\nb^\nu\widetilde N
  -\frac1{n-1} k^{\mu\nu}\Box\widetilde N\right)+
\\
  &+\nb_\rho(N^\rho P^{\mu\nu})+\frac2{n-1} P^{\mu\nu}\nb_\rho N^\rho
  - P^{\mu\rho}\nb_\rho N^\nu- P^{\nu\rho}\nb_\rho N^\mu,
\end{align*}
where
\begin{equation*}
  N_\mu=N^\nu k_{\mu\nu},~~~~\nb^\mu=k^{\mu\nu}\nb_\nu,~~~~
  \Box=k^{\mu\nu}\nb_\mu \nb_\nu,
\end{equation*}
and
\begin{equation*}
\begin{split}
  \hat R_{\mu\nu}&=R^{(k)}_{\mu\nu}+\frac{n-3}{n-2}\frac{\pl^2_{\mu\nu}\rho}\rho
  -\frac{(n-1)(n-3)}{(n-2)^2}\frac{\pl_\mu\rho\pl_\nu\rho}{\rho^2}
  -\frac{n-3}{n-2}\G^{(k)}_{\mu\nu}{}^\s\frac{\pl_\s \rho}\rho-
\\
  &-\frac1{(n-2)^2}k^{\mu\nu}\frac{k^{\s\lm}\pl_\s\rho\pl_\lm\rho}{\rho^2}
  +\frac1{n-2}k_{\mu\nu}\frac{\pl_\s(k^{\s\lm}\pl_\lm\rho)}\rho.
\end{split}
\end{equation*}

We now consider the constraints algebra. It is changed because we introduced
the new constraint $K_\bot$ instead of the dynamical constraint $H_\bot$.
Simple calculations yield
\begin{align}                                           \label{enewcz}
  [K_\bot,K'_\bot]&=-(\rho^2H_\mu k^{\mu\nu}
  +\rho^{\prime 2}H'_\mu k^{\prime\mu\nu})\dl_\nu,
\\                                                      \label{eneseo}
  [K_\bot,H'_\mu]&=-(K_\bot+K'_\bot)\dl_\mu,
\\                                                      \label{enekip}
  [H_\mu,H'_\nu]&=-H_\nu\dl_\mu-H'_\mu\dl_\nu.
\end{align}
Changes occur in Poisson brackets (\ref{enewcz}) and (\ref{eneseo})
as compared with the original algebra (\ref{econoo})--(\ref{econom}).
The second Poisson bracket, bracket (\ref{eneseo}), has a kinematical origin
and is defined by the fact that the new constraint is not a scalar function
but a tensor density of degree $\deg K_\bot=-2$. Poisson bracket (\ref{enewcz})
from direct calculations. We note that calculating that bracket is much simpler
in the new variables than in the original ones.

So far we have considered the metric density $k_{\mu\nu}$ and its conjugate
momenta $P^{\mu\nu}$ together with additional constraints (\ref{econev}).
This is possible in the classic theory, but problems arise in the quantum
theory of gravity. In the functional integral, considered in the next section,
we integrate over all values of $k_{\mu\nu}$ and $P^{\mu\nu}$. In principle,
we can solve the constraints explicitly, but doing so is not interesting,
because we then no longer have polynomials. Therefore, we describe the manifold
$\MN$ given by the coordinates $k_{\mu\nu}$ and $P^{\mu\nu}$ in detail. For
simplicity, we assume that the coordinates take all possible real values, and
the manifold $\MN$ is consequently topologically trivial and diffeomorphic to
the Euclidean space $\MR^{n(n-1)}$. We have the coordinate Poisson
brackets (\ref{eponek}) and (\ref{elpopp}) on that manifold, thus defining
the Poisson structure. It can be easily verified that this structure is
degenerate. This means that the manifold $\MN$ is not a symplectic and is only
a Poisson manifold (see, i.e., \cite{ChoDeW89}). Because the rank of the
Poisson structure is $n(n-1)-2$, there are two functionally independent
Casimir functions on the Poisson manifold $\MN$,
\begin{equation}                                        \label{ecasfu}
  C_1=\det k_{\mu\nu},~~~~C_2=\frac1{n-1}P^{\mu\nu}k_{\mu\nu}
\end{equation}
(introducing the constant factor $1/(n-1)$ in $C_2$ simplifies several
formulas in what follows.) Indeed, the Poisson brackets of the functions
$C_1$ and $C_2$ with all coordinates are zero,
\begin{equation*}
  [C_{1,2},k'_{\mu\nu}]=[C_{1,2},P^{\prime\mu\nu}]=[C_{1,2},\rho']=[C_{1,2},P']=0
\end{equation*}
as a consequence of the definition of Poisson structure (\ref{eponek}),
(\ref{elpopp}). The Poisson brackets of these functions with an arbitrary
differentiable function $f\in\CC^1(\MN)$ then vanish, $[C_{1,2},f']=0$, and
$C_1$ and $C_2$ are therefore Casimir functions. The Poisson structure
projected on sections $\MV\subset\MN$ defined by the equations $C_{1,2}=\const$
is nondegenerate. These sections are hence symplectic.

It is always possible to choose local coordinates on the Poisson manifold
$\MN$ that are connected with the symplectic leaves $C_{1,2}=\const$. We let
$(q_\Sa,p^\Sa)$, $\Sa,\Sb,\dotsc=1,\dotsc,n(n-1)/2-1$, denote the coordinates
on these leaves. We choose coordinates $q_\Sa$ and
$p^\Sa$ such that the representation
\begin{equation}                                        \label{epredk}
  k_{\mu\nu}=|C_1|^{\frac1{n-1}}\overset{\circ}k_{\mu\nu},~~~~
  P^{\mu\nu}=\overset{\circ}P{}^{\mu\nu}+C_2k^{\mu\nu},
\end{equation}
is satisfied, where the respective matrix elements $\overset{\circ}k_{\mu\nu}(q)$
and $\overset{\circ}P{}^{\mu\nu}(p)$ depend only on $q_\Sa$ and $p^\Sa$.
Obviously, such representations always exist. We choose the Casimir functions
themselves as the lacking coordinates on $\MN$. We thus obtain the local
coordinate system
\begin{equation}                                        \label{etracp}
  (k_{\mu\nu},P^{\mu\nu})~~\leftrightarrow~~(q_\Sa,C_1),(p^\Sa,C_2).
\end{equation}
Constraints (\ref{econev}) have a simple form in the new coordinates,
\begin{equation}                                        \label{ekazfc}
  C_1=\pm 1,~~~~C_2=0.
\end{equation}
We choose the plus or minus sign for $C_1$ if the space has the respective
even or odd dimensionality.

We consider the canonical transformation (\ref{eoldme})--(\ref{eoldmt})
from a different standpoint. Strictly speaking, a canonical transformation
considered in this section is canonical only between coordinates
\begin{equation}                                        \label{ecatrq}
  (g_{\mu\nu},p^{\mu\nu})~~\leftrightarrow~~(\rho,P),(q_\Sa,p^\Sa).
\end{equation}
The new phase space of general relativity in the considered case is the manifold
$\MR_+\times\MR\times\MV$, where $\rho\in\MR_+$, $P\in\MR$, and the submanifold
$\MV\subset\MN$ is defined by two values of Casimir functions (\ref{ekazfc}).
The Poisson brackets on $\MV$ have the canonical form by construction,
\begin{equation*}
  [q_\Sa,p^\Sb]=\dl_\Sa^\Sb,~~~~~[q_\Sa,q_\Sb]=0,~~[p^\Sa,p^\Sb]=0.
\end{equation*}
The constraints are made polynomial by extending the space $\MV$ to the Poisson
manifold $\MN$ with Poisson brackets (\ref{eponer})--(\ref{elpopp}). When
additional constraints (\ref{econev}) are solved explicitly, the polynomiality
is lost. This is not surprising. For example, electrodynamics contains
constraints whose explicit solution even leads to a nonlocal action for
physical degrees of freedom (see, i.e., \cite{GitTyu86}).

The Poisson brackets of the Casimir functions between themselves equal
zero $[C_1,C_2]=0$. From the standpoint of the Hamiltonian formalism, they
could be regarded as the first-class constraints generating gauge
transformations. But these transformations are trivial because the Poisson
brackets of Casimir functions with all phase-space coordinates vanish. This
is possible only on a Poisson manifold with a degenerate Poisson structure.
There are no Casimir functions on a symplectic manifold.

The Poisson manifold $\MN$ can be equipped with the second, now canonical,
Poisson bracket. With respect to this new canonical Poisson structure,
the submanifold $\MV$ is defined by two second-class constraints (\ref{ekazfc}).
Then the original degenerate Poisson structure (\ref{eponek}), (\ref{elpopp})
is just the Dirac bracket with respect to the canonical Poisson structure on $\MN$.

We also note the following. In gravity, we assume that a space-time metric
$g_{\al\bt}$ has the Lorentzian signature and therefore is nondegenerate.
In quantum gravity, the functional integral is integrated over all
independent metric components, and this property can be taken into account
only by restricting the integration domain. When the phase space is extended
to a Poisson manifold, the integration domain is extended to
the Euclidean space. The nondegeneracy of the metric is automatically provided
by the presence of $\dl$-functions in the integrand.
\section{Generating functional for the Green functions    \label{sgenfu}}
Here, we summarize the results calculating in the most interesting case
of the four dimensional space-time and write the explicit expression for
the generating functional for Green's functions. The polynomial Hamiltonian
formulation of general relativity on the Poisson manifold $\MR_+\times\MR\times\MN$
with coordinates $(\rho, P)\in\MR_+\times\MR$ and $(k_{\mu\nu},P^{\mu\nu})\in\MN$
is given in the preceeding section. The dimensionality of this manifold for $n=4$
equals 14. The Poisson structure is defined by the nonzero Poisson brackets
\begin{align*}
  [\rho,P']&=\dl,
\\
  [k_{\mu\nu},P^{\prime\rho\s}]&=\left(\dl_{\mu\nu}^{\rho\s}
  -\frac13k_{\mu\nu}k^{\rho\s}\right)\dl,
\\
  [P^{\mu\nu},P^{\prime\rho\s}]&=\frac13(P^{\mu\nu}k^{\rho\s}
  -P^{\rho\s}k^{\mu\nu})\dl.
\end{align*}
It is degenerate and has rank 12, which coincides with the dimensionality of
the phase space of general relativity. There are two Casimir functions
(\ref{ecasfu}) on $\MN$. The section $C_1=-1$, $C_2=0$ is a symplectic
submanifold $\MV\in\MN$ and defines the phase space of general relativity.

The action of general relativity in the new coordinates has the form
\begin{equation}                                        \label{ehieik}
  S_{\Sh\Se}=\int dx(P\dot\rho+P^{\mu\nu}\dot k_{\mu\nu}-H-\pl_\mu B^\mu),
\end{equation}
where the Hamiltonian density $H$ is equal to a linear combinations of constraints,
\begin{equation*}
  H=\tilde N K_\bot+N^\mu H_\mu,
\end{equation*}
and $\tilde N$ and $N^\mu$ are Lagrange multipliers. To action (\ref{ehieik}),
we added the boundary term $\pl_\mu B^\mu$ on a space section $x^0=\const$,
which is written as the divergence of some function of the canonical variables
$B^\mu(\rho,k_{\mu\nu},P,P^{\mu\nu})$. We briefly discuss the necessity of adding
this important term in the action below without specifying its form. The constraints
\begin{align}                                           \label{efirco}
  K_\bot&=P^{\mu\nu}P_{\mu\nu}-\frac16\rho^2P^2-\rho^2R^{(k)}
  -2\rho\pl_\mu(k^{\mu\nu}\pl_\nu\rho)+\frac32k^{\mu\nu}\pl_\mu\rho\pl_\nu\rho,
\\                                                      \label{esecon}
  H_\mu&=-2\pl_\nu(P^{\nu\s}k_{\s\mu})+P^{\nu\s}\pl_\mu k_{\nu\s}
  -\frac23\pl_\mu(P\rho)+P\pl_\mu\rho,
\end{align}
are polynomial first-class constraints and satisfy the algebra
(\ref{enewcz})--(\ref{enekip}). The scalar curvature $R^{(k)}$ for the metric
density $k_{\mu\nu}$ with unit determinant has the form (\ref{escuka}).
The constraint $K_\bot$ is quadratic in the momenta and the variable $\rho$.
It is a fifth-order polynomial in metric density $k_{\mu\nu}$ (and its
partial derivatives). The constraint $H_\mu$ is linear in the momenta and
also in the coordinates.

The expression for the generating functional for Green's functions as a
functional integral over the phase space \cite{Faddee69} is easily generalized
on a Poisson manifold. For brevity, we introduce a new notation for the secondary
constraints and Lagrange multipliers:
\begin{equation*}
  \lbrace H_a\rbrace=\lbrace K_\bot,H_\mu\rbrace,~~~~
  \lbrace N^a\rbrace=\lbrace \tilde N,N^\mu\rbrace,~~~~
  a=0,1,2,3.
\end{equation*}
We now fix the invariance under general coordinate transformations
using four gauge conditions $F^a=0$. The gauge is assumed to be canonical,
\begin{equation*}
  \det[H_a,F^b]\ne0.
\end{equation*}

The canonical form of the generating functional for the Green's functions for
the metric is given by the functional integral up to a normalization factor
\cite{Faddee69}:
\begin{multline}                                       \label{efunor}
  Z(J)=\int D(g_{\mu\nu})D(p^{\mu\nu})D(N^a)
  \exp\left\lbrace\frac i h \int dx(p^{\mu\nu}\dot g_{\mu\nu}
  -N^aH_a-\pl_\mu B^\mu+g_{\mu\nu}J^{\mu\nu})\right\rbrace\times
\\
  \times\det[H_a,F^b]\prod_a\dl(H_a)\prod_a\dl(F^a),
\end{multline}
where
\begin{equation*}
  D(g_{\mu\nu})=\prod_{x,~\mu\le\nu}dg_{\mu\nu},~~~~
  D(p^{\mu\nu})=\prod_{x,~\mu\le\nu}dp^{\mu\nu},~~~~
  D(N^a)=\prod_{x,~a}dN^a
\end{equation*}
and the $J^{\mu\nu}$ are the sources for the metric. The functional integral
$Z(J)$ is the generating functional only for ``coordinate'' Green's functions
because the sources are written only for the space metric \cite{GitTyu86}.
Because the Jacobian of any canonical transformation and, in particular, of
transformation (\ref{ecatrq}) is equal to unity, the expression for the
functional integral can be rewritten in the equivalent form
\begin{equation*}
\begin{split}
  Z(J)=&\int D(\rho) D(P)D(q_\Sa) D(p^\Sa) D(N^a)\times
\\
  &\times\exp\left\lbrace\frac i h \int dx(P\dot\rho+p^\Sa\dot q_\Sa
  -N^aH_a-\pl_\mu B^\mu+\rho J+q_\Sa J^\Sa)\right\rbrace\times
\\
  &\times\det[H_a,F^b]\prod_a\dl(H_a)\prod_a\dl(F^a).
\end{split}
\end{equation*}

The constraints $H_a$ are nonpolynomial in coordinates $q_\Sa$ and momenta
$p^\Sa$ in this form. The integration must be extended over the whole
Poisson manifold $\MN$ to make the constraints polynomial. We provide this
by introducing two additional $\dl$-functions:
\begin{equation}                                           \label{efuncf}
\begin{split}
  Z(J)=&\int D(\rho)D(P)D(k_{\mu\nu})D(P^{\mu\nu})D(N^a)\times
\\
  &\times\exp\left\lbrace\frac i h \int dx(P\dot\rho+P^{\mu\nu}\dot k_{\mu\nu}
  -N^aH_a-\pl_\mu B^\mu+\rho J+k_{\mu\nu}\tilde J^{\mu\nu})\right\rbrace\times
\\
  &\times\det[H_a,F^b]\dl(C_1+1)\dl(C_2)\prod_a\dl(H_a)\prod_a\dl(F^a).
\end{split}
\end{equation}
Two new $\dl$-functions remove the integration over the additional variables
and restrict the integration over the Poisson manifold $\MN$ to integration
over symplectic section (\ref{ekazfc}). To perform the corresponding integration,
we must perform coordinates transformation (\ref{etracp}).

We must prove that the Jacobian of the coordinates transformation (\ref{etracp})
\begin{equation*}
  D(q_\Sa)D(C_1)D(p^\Sa)D(C_2)=D(k_{\mu\nu})D(P^{\mu\nu})
  \left|\frac{\pl(q_\Sa,C_1,p^\Sa,C_2)}{\pl(k_{\mu\nu},P^{\mu\nu})}\right|
\end{equation*}
equals unity on constraint surface (\ref{ekazfc}) to prove that the proposed
expression for generating functional (\ref{efuncf}) over the Poisson manifold
is equivalent to the original functional integral (\ref{efunor}) over the
phase space.This can be easily done. Let the coordinates $q_\Sa$ parametrize
the matrix $k_{\mu\nu}$ with a unit determinant arbitrarily. Then
\begin{equation*}
  \left|\frac{\pl(q_\Sa,C_1,p^\Sa,C_2)}{\pl(k_{\mu\nu},P^{\mu\nu})}\right|
  =\left|\frac{\pl(q_\Sa,C_1)}{\pl(k_{\mu\nu})}\right|
  \left|\frac{\pl(p^\Sa,C_2)}{\pl(P^{\mu\nu})}\right|,
\end{equation*}
because the matrix
\begin{equation*}
  \frac{\pl(q_\Sa,C_1)}{\pl(P^{\mu\nu})}=0.
\end{equation*}
As a consequence of the definition of canonical momenta (\ref{eoldmo}), we obtain
\begin{equation}                                           \label{expnep}
  p^\Sa=-\frac{\dl F}{\dl q_\Sa}=-\frac{\dl F}{\dl k_{\mu\nu}}
  \frac{\pl k_{\mu\nu}}{\pl q_\Sa}=P^{\mu\nu}\frac{\pl k_{\mu\nu}}{\pl q_\Sa}.
\end{equation}
Representation (\ref{epredk}) for the metric $k_{\mu\nu}$ yields
\begin{equation*}
  k_{\mu\nu}=(n-1)C_1\frac{\pl k_{\mu\nu}}{\pl C_1}.
\end{equation*}
Therefore,
\begin{equation*}
  \frac{\pl C_2}{\pl P^{\mu\nu}}=\frac1{n-1}k_{\mu\nu}
  =C_1\frac{\pl k_{\mu\nu}}{\pl C_1}.
\end{equation*}
Consequently,
\begin{equation*}
  \frac{\pl(p^\Sa,C_2)}{\pl(P^{\mu\nu})}=\frac{\pl k_{\mu\nu}}{\pl(q_\Sa,C_1)}C_1.
\end{equation*}
This implies that for the constraint surface, the modulus of the Jacobian of
the coordinates transformation is equal to unity:
\begin{equation*}
  \left|\frac{\pl(q_\Sa,C_1,p^\Sa,C_2)}{\pl(k_{\mu\nu},P^{\mu\nu})}\right|
  _{C_1=-1}=1.
\end{equation*}
This motivates introducing the numerical factor in the Casimir function
$C_2$ in Eq.\ (\ref{ecasfu}). Without it, the Jacobian of the coordinate
transformation would equal some nonzero constant, which could be included
in the definition of the normalization factor of the generating functional.

We now say a few words about including the boundary term $\pl_\mu B^\mu$
in the action of general relativity. The original expression for the generating
functional (\ref{efunor}) is justified by assuming that the functional integral
over just physical degrees of freedom with the unit measure appears after
solution of all the constraints and gauge conditions. The Hamiltonian on the
constraint surface becomes zero in this case, $N^a H_a=0$. On the other hand,
we know that the dynamics of the physical degrees of freedom are nontrivial.
A possible way out from this contradiction was proposed in \cite{ArDeMi62}.
If the gauge condition depends explicitly on time, then the nontrivial
Hamiltonian for the physical degrees of freedom on the constraint surface
arises from the kinetic term $p^{\mu\nu}\dot g_{\mu\nu}$. But this is not a
unique possibility. To obtain a nontrivial Hamiltonian for the physical degrees
of freedom in the canonical gauge, which does not depend on time explicitly,
the boundary term was added to the action \cite{GitTyu86}. The importance of
the boundary term in general relativity is now universally recognized, but
its role still remains obscure in many cases because of large technical
difficulties. In two-dimensional gravity where the constraints are solved
explicitly, it was proved that the Hamiltonian for the physical degrees
of freedom appears from the boundary term for gauge conditions, which do not
depend explicitly \cite{Katana02}. Namely, the term $\pl_\mu B^\mu$
on the constraint surface is equal to the Hamiltonian density for the physical
degrees of freedom and is not equal to the divergence of some function.
This statement is local and is independent of whether the universe is closed.

We note one more important point. For an asymptotically flat space-time, the
volume integral of $\pl_\mu B^\mu$ is equal to the surface integral of
$B^\mu$ and coincides with the mass of the Schwarzschild solution.
That is why this integral was proposed as the definition of the total energy
of a gravitational field \cite{ArDeMi62}. It is usually assumed that the total
energy of a closed universe vanishes because it has no boundary, and no surface
integral arises. This is true if the constraints admit smooth solutions on compact
manifolds for the nonphysical degrees of freedom in the general case. The example
of two-dimensional gravity (which includes spherically symmetric solutions of
general relativity) shows that constraints do not admit smooth solutions on a
circle in general. In this case, we must make a cut on a compact space and add
a boundary term there in order to pose the variational problem. This yields
a nontrivial expression for the total energy of a closed universes.
We can say this in other words. The divergence $\pl_\mu B^\mu$ on the constraint
surface expressed in terms of the physical degrees of freedom is no longer the
divergence of some function. This Hamiltonian density for the physical degrees
of freedom has the same form independently of whether the universe is closed or
not. Therefore, this expression can be taken as the definition of the energy
density of a gravitational field. Of course, this definition is not covariant
and depends on the choice of a coordinate system.
\section{Homogeneous and isotropic universe           \label{shoisp}}
We consider the Friedman universe \cite{Friedm22,Friedm24} to
demonstrate the new variables introduced in Sec.\ \ref{scatrg}. First,
we recall the derivation of the equations in the Lagrangian formalism.
For comparison, we then reformulate the model in the Hamiltonian language
in the old and new variables in which the Hamiltonian becomes polynomial.

In the Friedman model of the universe, which is the basis for most
contemporary cosmological models, we assume that the space is a
Riemannian manifold of constant curvature at each instant (all sections
$x^0=t=\const$). This assumption corresponds to a homogeneous isotropic
universe. The metric satisfying this requirement is
\begin{equation}                                        \label{efriem}
  ds^2=dt^2+a^2\overset{\circ}g_{\mu\nu}dx^\mu dx^\nu,
\end{equation}
where $a=a(t)$ is the scale factor depending only on time, and
$\overset{\circ}g_{\mu\nu}$ is a (negative-definite) constant-curvature
space metric, which is independent of time by assumption. The specific
form of the constant-curvature space metric $\overset{\circ}g_{\mu\nu}$ depends
on the coordinate system and is not important for the following consideration.
For simplicity, we restrict ourselves to the most important case of
four-dimensional space-time ($n=4$).

The metric of the space-time satisfies Einstein's equations
\begin{equation}                                        \label{eincos}
  R_{\al\bt}-\frac12Rg_{\al\bt}+\frac12\Lm g_{\al\bt}=-\frac12T_{\al\bt},
\end{equation}
where we introduce the cosmological constant $\Lm$ and the matter energy-momentum
tensor $T_{\al\bt}$. The energy-momentum tensor in the comoving frame
has the form \cite{LanLif88}
\begin{equation}                                             \label{engmom}
  T^\al{}_\bt=\begin{pmatrix} {\cal E} & 0 & 0 & 0 \\ 0 & -{\cal P} & 0 & 0 \\
  0 & 0 & -{\cal P} & 0 \\ 0 & 0 & 0 & -{\cal P} \end{pmatrix},
\end{equation}
where ${\cal E}$ and ${\cal P}$ are the respective energy density and pressure
of matter. For a homogeneous isotropic universe in the chosen coordinate
system, these densities depend only on time ${\cal E}={\cal E}(t)$ and
${\cal P}={\cal P}(t)$.

We also assume that the matter equation of state is given by some
dependence  of energy density on pressure,
\begin{equation}                                              \label{eqstun}
  {\cal E}={\cal E}({\cal P}).
\end{equation}

Because energy-momentum tensor (\ref{engmom}) was not obtained by
varying some invariant action for matter fields with respect to a metric,
the energy density ${\cal E}$ cannot be an arbitrary function. Indeed, the
covariant divergence of the left hand side of Einstein's equations is
identically zero as a consequence of the Bianchi identities. Therefore,
Einstein's equations yield the equation for the energy-momentum tensor:
\begin{equation*}
  \nb_\al T^\al{}_\bt=0.
\end{equation*}
For metric (\ref{efriem}) and energy-momentum tensor (\ref{engmom}), these
four relations reduce to one nontrivial equation
\begin{equation}                                        \label{enmoeq}
  \dot{\cal E}+\frac{3\dot a}a({\cal E}+{\cal P})=0.
\end{equation}
For a given equation of state (\ref{eqstun}), we have one first-order
differential equation, which we rewrite in the form
\begin{equation}                                        \label{eqstea}
  \frac{d{\cal E}}{{\cal E}+{\cal P}({\cal E})}=-3\frac{da}a.
\end{equation}
Solving this equation yields the energy density ${\cal E}$ as a
function of the scale factor $a$.

We obtain the equation for the scale factor because metric (\ref{efriem})
must satisfy Einstein's equations (\ref{eincos}). Simple calculations yield
the expressions for the Einstein tensor
\begin{equation*}
\begin{split}
  R_0{}^0-\frac12R&=3\frac{K_0-\dot a^2}{a^2},
\\
  R_0{}^\mu&=0,
\\
  R_\mu{}^\nu-\frac12R\dl_\mu^\nu&=-\frac1{a^2}\left(2a\ddot a+\dot a^2
  -K_0\right)\dl_\mu^\nu,
\end{split}
\end{equation*}
where $K_0$ in our notation is the scalar curvature of a three-dimensional
sphere $(K_0=1)$, Euclidean space $(K_0=0)$, or one-sheet hyperboloid $(K_0=-1)$.
We can now easily see that Einstein's equations lead to only two nontrivial
equations on the scale factor:
\begin{align}                                           \label{einsco}
  3\frac{K_0-\dot a^2}{a^2}+\Lm+\frac12{\cal E}&=0,
\\                                                      \label{einsct}
  -\frac1{a^2}\left(2a\ddot a+\dot a^2-K_0\right)+\Lm-\frac12{\cal P}&=0.
\end{align}
It is easy to verify that Eq.\ (\ref{einsct}) is a consequence of Eqs.\
(\ref{einsco}) and (\ref{enmoeq}) because Einstein's equations are linearly
dependent as soon as Eq.\ (\ref{enmoeq}) is satisfied. Therefore, Eq.\
(\ref{einsct}) can be dropped, but we do not do this, because it is needed
for the canonical treatment of the Friedman universe.

We thus find the dependence ${\cal E}={\cal E}(a)$ of the energy density on
the scale factor by solving Eq.\ (\ref{eqstea}) for a given equation of state
(\ref{eqstun}). Substituting this function in Eq.\ (\ref{einsco}), we obtain
the first-order ordinary differential equation for the scale factor. This is
precisely the main equation in the standard cosmological models for a
homogeneous isotropic universe.

We start with the canonical formulation. For metric (\ref{efriem}), we have
\begin{equation*}
  N=1,~~~~N_\mu=0,~~~~g_{\mu\nu}=a^2\overset{\circ}g_{\mu\nu},
\end{equation*}
External curvature tensor (\ref{excurv}) and the volume element are
\begin{equation*}
\begin{split}
  K_{\mu\nu}&=-a\dot a\overset{\circ}g_{\mu\nu},~~~~K=-3\frac{\dot a}a.
\\
  \hat e&=a^3\overset{\circ}e,~~~~
  \overset{\circ}e=\sqrt{|\det\overset{\circ}g_{\mu\nu}|}.
\end{split}
\end{equation*}
Canonical momenta (\ref{emomet}) conjugate to the space metric become
\begin{equation*}
  p^{\mu\nu}=-2\dot a\overset{\circ}e\overset{\circ}g^{\mu\nu}.
\end{equation*}
The traceless part of these momenta is identically zero,
\begin{equation}                                        \label{emomcm}
  \tilde p^{\mu\nu}=0,~~~~p=-6a^2\dot a\overset{\circ}e.
\end{equation}
The Hamiltonian density is given by the single dynamical constraint (\ref{ecoadm})
because the shift function for the Friedman universe is equal to zero
\begin{equation}                                         \label{edycoa}
  H_\bot=-6\overset{\circ}e\left[(\dot a^2-K_0)a
  -\frac13\Lm a^3-\frac16{\cal E}a^3\right],
\end{equation}
where we take the contribution from cosmological constant and matter fields
into account. We cannot insert the momentum trace $p$ instead of the time
derivative $\dot a$ in this expression, because it is not the variable
conjugate to the scale factor.

To find the momentum conjugate to $a$, we rewrite Lagrangian (\ref{eadmla})
for metric (\ref{efriem}),
\begin{equation*}
  \CL_{\Sa\Sd\Sm}=-6\overset{\circ}e\left[(\dot a^2+K_0)a
  +\frac13\Lm a^3+\frac16\CE a^3\right].
\end{equation*}
This expression can be integrated over the space because all the dependence
on the space coordinates is contained in the volume element $\overset{\circ}e$.
Dropping the constant factor $-6V$, where $V$ is the volume of space (infinite
for the Euclidean space and one sheet hyperboloid), we obtain the Lagrangian
for the scale factor
\begin{equation}                                   \label{elagma}
  L=(\dot a^2+K_0)a+\frac13\Lm a^3+\frac16\CE a^3,
\end{equation}
which is independent on the space coordinates. This is the standard Lagrangian
for a point particle moving in a one-dimensional space with the coordinate
$a\in\MR_+$.

Deriving the Lagrangian for the scale factor, we dropped the negative factor
to obtain the positive sign of the kinetic term $a\dot a^2$. We thus changed
the total sign of the action and hence the sign of the energy. We note that the
contribution of the kinetic term of the scale factor to the energy is negative.

The expression for the momentum conjugate to the scale factor follows from
Lagrangian (\ref{elagma}):
\begin{equation*}
  p_a=\frac{\pl L}{\pl\dot a}=2a\dot a,
\end{equation*}
which differs from momentum trace (\ref{emomcm}) by a factor and is independent
on the space coordinates. The Hamiltonian for the scale factor corresponding
to Lagrangian (\ref{elagma}) is
\begin{equation}                                     \label{ehaafa}
  H=\frac1{4a}p_a^2-K_0 a-\frac13\Lm a^3-\frac\CE6 a^3.
\end{equation}

We see that this expression for the Hamiltonian coincides up to the factor
$-6V$ with the expression obtained by integrating dynamical constraint
(\ref{edycoa}) over the space. This observation is nontrivial because
some equations of motion may be lost when specific expressions for field
variables are inserted into the Lagrangian.

The Hamiltonian equations for the scale factor are
\begin{equation}                                        \label{eqhafa}
\begin{split}
  \dot a&=\frac1{2a}p_a,
\\
  \dot p_a&=\frac1{4a^2}p_a^2+K_0+\Lm a^2+\frac12\CP a^2,
\end{split}
\end{equation}
where we use Eq. (\ref{eqstea}) to calculate the Poisson bracket
\begin{equation*}
  [p_a,\CE]=[p_a,a]\frac{d\CE}{da}=-3\frac{\CE+\CP}a.
\end{equation*}
It can be easily verified that Hamiltonian equations (\ref{eqhafa}) are
equivalent to the second-order Lagrangian equation (\ref{einsct}), and
Hamiltonian (\ref{ehaafa}) is proportional to the left-hand side of Eq.\
(\ref{einsco}).

We have thus formulated equations for the Friedman universe in the
Hamiltonian language. In contrast to the Hamiltonian particle dynamics, we
have an additional constraint along with canonical equations of motion
(\ref{eqhafa}),
\begin{equation}                                       \label{ehacoa}
  H(a,p_a)=0.
\end{equation}
In other words, we only seek those solutions of the equations of motion
for which the energy is equal to zero. This problem is self-consistent because
the energy is conserved. We have thus formulated equations (\ref{einsco}) and
(\ref{einsct}) for the scale factor in the Hamiltonian form and proved that
one constraint (\ref{ehacoa}) on the scale factor and the corresponding momentum
is lost if the expression for the metric (\ref{efriem}) is substituted not in
Einstein's equations but in the action.

Hamiltonian (\ref{ehaafa}) and equations of motion (\ref{eqhafa}) are
nonpolynomial in the scale factor. We show what happens with the equations
under the canonical transformation described in section \ref{scatrg}. As a
consequence of Eqs.\ (\ref{eoldme})--(\ref{eoldmt}), we obtain expressions
for the canonical variables
\begin{align}                                           \label{exprka}
  \rho&=~a^2\overset{\circ}e{}^{2/3},~~~~k_{\mu\nu}=\eta_{\mu\nu},
\\                                                      \label{expcpp}
  P&=-2\frac{\dot\rho}{\sqrt\rho},~~~~P^{\mu\nu}=0,
\end{align}
external curvature
\begin{equation*}
  K_{\mu\nu}=-\frac12\dot\rho\eta_{\mu\nu},~~~~K=-\frac{3\dot\rho}{2\rho}
\end{equation*}
and volume element
\begin{equation*}
  \hat e=\rho^{3/2}
\end{equation*}
after the canonical transformation.

To simplify calculations, it is easiest to separate variables by extracting
the factor $q(t)$ depending only on time from $\rho$,
\begin{equation*}
  \rho(t,x^\mu)=q(t)\overset{\circ}e^{2/3}.
\end{equation*}
The Lagrangian density can then be integrated over the space as previously,
and the Hamiltonian reformulation of the equations reduces to the
redefinition of the scale factor
\begin{equation}                                        \label{escaha}
  q=a^2.
\end{equation}
Lagrangian (\ref{elagma}) for the new variable is
\begin{equation*}
  L=\frac1{4\sqrt q}\dot q^2+K_0\sqrt q+\frac13\Lm q^{3/2}+\frac16\CE q^{3/2}.
\end{equation*}
The momentum conjugate to the new dynamical variable $q(t)$ is
\begin{equation*}
  p_q=\frac1{2\sqrt q}\dot q.
\end{equation*}
The corresponding Hamiltonian contains a nonpolynomial factor,
\begin{equation*}
  H=\sqrt q\left(p_q^2-K_0-\frac13\Lm q-\frac16\CE q\right).
\end{equation*}
During the construction of the polynomial Hamiltonian formulation, the
dynamical constraint was multiplied by factor (\ref{etrakh}). As a result,
we obtain a new constraint, which is now polynomial, and the new Hamiltonian
\begin{equation}                                         \label{enehak}
  K=q^{3/2}H=q^2\left(p_q^2-K_0-\frac13\Lm q-\frac16\CE q\right).
\end{equation}
To preserve the Hamiltonian form of the equations of motion we must also
redefine time as well $t\rightarrow \tau$, where the new parameter is
defined by the differential equation
\begin{equation*}
  \frac{d\tau}{dt}=q^{-3/2}=a^{-3}.
\end{equation*}
Redefining time corresponds to redefining Lagrange multiplier (\ref{elagtr}).
The equations of motion for Hamiltonian (\ref{enehak}) are
\begin{equation*}
\begin{split}
  \frac{dq}{d\tau}&=2q^2p_q,
\\
  \frac{dp_q}{d\tau}&=q^2\left(\frac13\Lm-\frac1{12}\CE-\frac14\CP\right)
  -\frac2q K,
\end{split}
\end{equation*}
where we once again used Eq.\ (\ref{eqstea}). The second term in the second
equation can be dropped because the model contains the constraint
\begin{equation}                                        \label{econkz}
  K(q,p_q)=0.
\end{equation}
Thus in the new variables, we have a ``particle'' described by the coordinate
$q(\tau)\in\MR_+$ and momentum $p_q(\tau)\in\MR$ with Hamiltonian (\ref{enehak})
and with constraint (\ref{econkz}) imposed on the canonical variables. The
Hamiltonian and equations of motion are polynomial (if the energy density
$\CE(q)$ is polynomial in $q$) and equivalent to the original
Einstein's equations for the scale factor (\ref{einsco}), (\ref{einsct}).

For the homogeneous isotropic universe we can go even further and
eliminate the common factor $q^2$ from Hamiltonian (\ref{enehak})
by redefining time $t\rightarrow \tau'$, where
\begin{equation*}
  d\tau'=q^{-1/2}dt.
\end{equation*}
We then have a ``particle'' with the simple Hamiltonian
\begin{equation*}
  K'=p_q^2-K_0-\frac13\Lm q-\frac16\CE q.
\end{equation*}
and the corresponding equations of motion
\begin{align}                                          \label{emocoq}
  \frac{dq}{d\tau'}&=2p_q,
\\                                                      \nonumber
  \frac{dp_q}{d\tau'}&=\frac13\Lm-\frac1{12}\CE-\frac14\CP.
\end{align}
The space curvature $K_0$ now does not contribute to the equations of
motion at all. The system of equations of motion is solved under the zero
constraint on the Hamiltonian, $K'=0$. Just as for the original system of
equations (\ref{einsco}) and (\ref{einsct}), the Hamiltonian equations of
motion in the Lagrangian form follow from the equation $K'=0$, where instead of
the momentum, we must substitute its expression in terms of the time
derivative of the scale factor (\ref{emocoq}). This equation is equivalent
to Eq.\ (\ref{einsco}).

For dust matter, we have $\CP=0$, and Eq.\ (\ref{eqstea}) can be easily integrated,
\begin{equation*}
  \CE=\frac M{a^3},~~~~M=\const,
\end{equation*}
which provides a massless particle moving in the potential
\begin{equation*}
  -\frac13\Lm q-\frac M{6q^2}.
\end{equation*}

Introducing new canonical variables thus allows reformulating the
equations for the Friedman universe in polynomial form. Because Eq.\
(\ref{einsco}) has been well studied in recent decades, the new formulation
is unlikely to yield new results for the analysis of classical solutions
in the considered case. It may be useful for constructing a quantum
model of the Friedman universe, but this question requires a separate
investigation and is beyond the scope of this paper. Our consideration here
is intended only to illustrate how the general method works in a simple case
where all steps can be verified by simple calculations.
\section{Conclusion}
We have demonstrated that considering the determinant of the metric and
the conjugate momentum as independent additional variables leads to a
Hamiltonian formulation of general relativity in which all the constraints
are polynomial. The model is formulated in polynomial form not in the phase
space but on a Poisson manifold where the Poisson bracket is degenerate. We
stress that the resulting model is equivalent to general relativity.

In the new variables, the canonical momenta are proportional to the
irreducible components of the momenta in the standard metric formulation.
This property essentially simplifies the calculations, in particular,
the calculations of the Poisson bracket of the dynamical constraint.

The proposed canonical formulation of general relativity allows writing
the functional integral on the Poisson manifold. We proved that this integral
is equivalent to the functional integral over the phase space. The advantage
of the new expression for the generating functional for the Green's functions
is that the action and all arguments of  $\dl$-functions are polynomial
in independent variables. This leads to the presence of only a finite number
of vertices in the diagram techniques. This seems to simplify calculations
in the quantum theory of gravity.

As an example of using new variables, we considered the Friedman model of the
universe. In these variables, the Hamiltonian and the equations of motion for
the scale factor are simplified and take a polynomial form.

A similar transformation of variables in the configuration space leading to
a polynomial Hilbert--Einstein action was proposed in \cite{Peres63} and
recently rediscovered \cite{Katana05B}.

The author is sincerely grateful to I.~V.~Volovich for the discussion of the
paper. This work is supported by Russian Foundation of Basic Research
(Grant No.\ 05-01-00884) and the Program for Supporting Leading
Scientific Schools (Grant No.\ NSh-6705.2006.1).


\end{document}